\documentclass[aps,prl,reprint,twocolumn,showpacs,preprintnumbers,letterpaper]{revtex4}
\usepackage{amssymb}
\usepackage{color}
\usepackage{graphicx}
\usepackage{appendix}
\usepackage{natbib}
\usepackage{float}
\usepackage{hyperref}
\usepackage{amsmath}
\usepackage[tight,FIGBOTCAP]{subfigure}
\usepackage[matrix,frame,arrow]{xypic}

\input{Qcircuit}
\begin{document}

\title{Ultrafast and Fault-Tolerant Quantum Communication across Long
Distances}
\author{Sreraman Muralidharan$^{1}$}
\author{Jungsang Kim$^{2}$}
\author{Norbert L\"utkenhaus$^{3}$}
\author{Mikhail D. Lukin$^{4}$}
\author{Liang Jiang$^{5}$}
\affiliation{$^1$Department of Electrical Engineering, Yale University, New Haven, CT
06511 USA}
\affiliation{$^2$Department of Electrical and Computer Engineering, Duke University,
Durham, NC 27708 USA}
\affiliation{$^3$Institute of Quantum computing, University of Waterloo, N2L 3G1
Waterloo, Canada}
\affiliation{$^4$Department of Physics, Harvard University, Cambridge, MA 02138, USA}
\affiliation{$^5$Department of Applied Physics, Yale University, New Haven, CT 06511 USA}
\date{\today }

\begin{abstract}
Quantum repeaters (QRs) provide a way of enabling long distance quantum
communication by establishing entangled qubits between remote locations. In
this Letter, we investigate a new approach to QRs in which quantum
information can be faithfully transmitted via a noisy channel without the
use of long distance teleportation, thus eliminating the need to establish
remote entangled links. Our approach makes use of small encoding blocks to
fault-tolerantly correct both operational and photon loss errors. We
describe a way to optimize the resource requirement for these QRs with the
aim of the generation of a secure key. Numerical calculations
indicate that the number of quantum memory bits at each repeater station required for the
generation of one secure key has favorable poly-logarithmic scaling with the
distance across which the communication is desired.
\end{abstract}

\maketitle

%\title{Ultrafast and fault-tolerant quantum communication across long distances}

Quantum communication across long distances ($10^{3}$-$10^{4}$km) can
significantly extend the applications of quantum information protocols such
as quantum cryptography \cite{Qcryp} and quantum secret sharing \cite%
{Qsecret1, Qsecret2} which can be used for the creation of a secure quantum
internet \cite{Qinternet}. Quantum communication can be carried out by first
establishing a remote entangled pair between the sender and the receiver and
using teleportation to transmit information faithfully. However, there are
two main challenges that have to be overcome. First, fiber attenuation
during transmission leads to an exponential decrease in entangled pair
generation rate. Second, several operational errors such as channel errors,
gate errors, measurement errors and quantum memory errors severely degrade
the quality of entanglement used for secure key generation. In addition,
quantum states cannot be amplified or duplicated deterministically in
contrast to classical information \cite{nocloning}. Establishing quantum
repeater (QR) stations based on entanglement distribution is the only
currently known approach to long-distance quantum communication using
conventional optical fibers without exponential penalty in time and
resources.

A number of schemes have been proposed for long distance quantum
communication using QRs \cite{Qrepeater, Qrepeater2, Qcomm, Coherent, Liang,
Fowler, Munro}, most of which could be broadly classified into three
classes. The first class of QRs \cite{Qrepeater,Qrepeater2,Qcomm,Coherent}
reduces the exponential scaling of fiber loss to polynomial scaling by
introducing intermediate QR nodes. However, this scheme for long distance
quantum communication is relatively slow \cite{Sangouard11}, even after
optimization \cite{JTKL07}, limited by the time associated with two-way
classical communication between remote stations required for the
entanglement purification process needed to correct operational errors \cite%
{Purification}. %Second Gen QR%
In contrast, the second class of QRs introduce quantum encoding and
classical error correction to replace the entanglement purification with
classical error correction, handling all operational errors \cite%
{Liang,Munro10}. As a consequence, the entanglement generation
rate further improves from $1/O(poly(L_{tot}))$ to $1/O(poly(log(L_{tot})))$
where $L_{tot}$ is the total distance of communication.
Recently, the approach to the third type of QRs was proposed, which uses
quantum encoding to deterministically correct photon losses\cite%
{Fowler,Munro}. By entirely eliminating two-way classical communication
between \emph{all} repeater stations, the third class of QRs promise
extremely high key generation rates that can be close to classical
communication rates, limited only by the speed of local operations.

\begin{figure}[h]
\includegraphics[width=8cm]{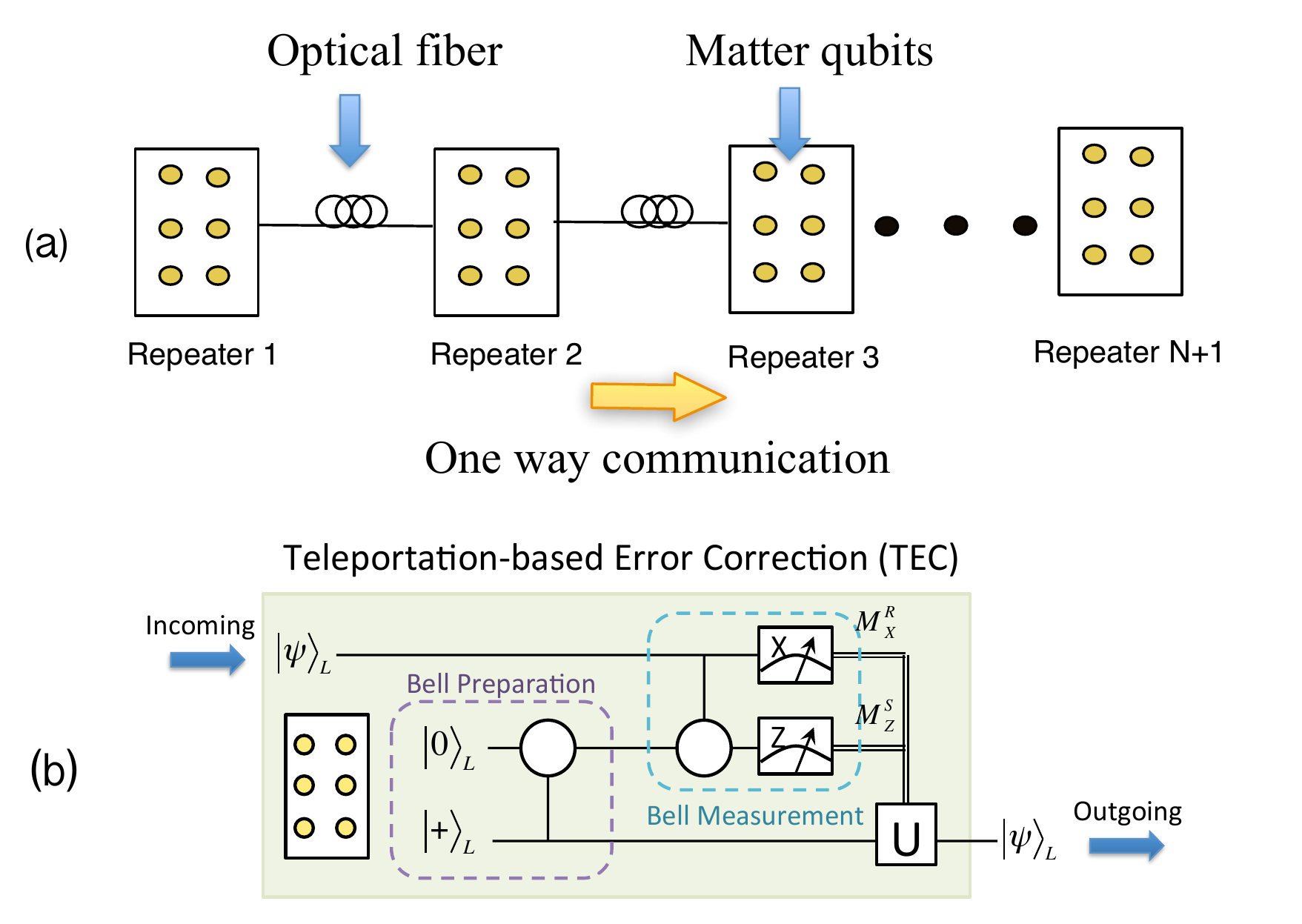}
\caption[fig:repeater]{(color online). (a) A schematic view of the third
class of QRs showing individual matter qubits in the repeater stations
connected by an optical fiber. The quantum state is encoded into an error
correcting code with photonic qubits, which are multiplexed and transmitted
through the optical fiber to the neighboring repeater station. The quantum
state of photonic qubits is transferred to the matter qubits and error
correction is performed. After the error correction procedure, the quantum
state of the matter qubits is transferred to photonic qubits and transmitted
to the next repeater station. This procedure is carried out until the
information reaches the receiver. (b) The TEC procedure consists of Bell
state preparation and Bell measurement at the encoded level. Each line in
the circuit represents an encoding block and the CNOT gate has a transversal
implementation for CSS codes. This TEC scheme can be potentially implemented
in a cavity QED system \protect\cite{Duan04b,Supp}.}
\label{fig:repeater}
\end{figure}

Besides high key generation rate, it is very important to consider the
resource requirement and fault-tolerant implementation for this type of QRs.
In the fault-tolerant surface-code proposal by Fowler \textit{et. al.} \cite%
{Fowler}, the resource for each station was estimated to scale
logarithmically with the distance, while the exact resource overhead was
found to be sensitive to the parameters for various imperfections. The
proposal by Munro \textit{et. al.} \cite{Munro} focused on the correction of
photon loss errors using quantum parity code (QPC) \cite{Shor}, but did not
consider fault tolerance, as perfect gate operations were assumed in their
analysis. %%% ----- Summarize Results in this Letter ------------%%%
In this Letter, we propose a fault-tolerant architecture for third class of
QRs, where a teleportation-based error correction (TEC) protocol \cite{Tele,
KLM} is employed \emph{within} each repeater station to correct both
operational and photon loss errors using Calderbank-Shor-Steane (CSS)
encoding.
%\footnote{The teleportation based error correction occurs within the repeater station
%and does not require two-way classical communication.}.
We quantitatively investigate the optimum resource requirements using a
\emph{cost function} and optimize it for different repeater parameters. A
schematic view of the proposed architecture of the third class of QRs is
presented in Fig.~\ref{fig:repeater}.

\paragraph*{Fault-tolerant architecture.}

Analogous to fault-tolerant quantum computers \cite{Nielsen}, fault-tolerant
QRs should reliably relay quantum information from one repeater station to
another in the presence of various imperfections. Unlike quantum computers,
QRs do not require a universal gate set and it is sufficient to have CNOT
gates and state initialization/measurement associated with the complimentary
basis of $\{|0\rangle ,|1\rangle \}$ and $\{|+\rangle ,|-\rangle \}$.
However, QRs are confronted by an important challenge from transmission
loss, which is less severe in most models of quantum computation. To design
fault-tolerant third class of QRs, we consider the CSS codes for their
fault-tolerant properties \cite{Nielsen}, in particular the compatibility
with the TEC protocol that can efficiently correct not only operational
errors, but also photon loss errors \cite{Tele, KLM}. The $\left( n,m\right)
$-QPC \cite{Shor} is a class of CSS codes with encoded qubits $|0\rangle
_{L}=\frac{1}{\sqrt{2}}(|+\rangle _{L}+|-\rangle _{L})$ and $|1\rangle _{L}=%
\frac{1}{\sqrt{2}}(|+\rangle _{L}-|-\rangle _{L})$, where $|\pm \rangle _{L}$
are given by
\begin{eqnarray}
&|+\rangle _{L}=&\frac{1}{2^{n/2}}{(|00...0\rangle _{12...m}+|11...1\rangle
_{12...m})}^{\otimes n}  \nonumber \\
&|-\rangle _{L}=&\frac{1}{2^{n/2}}{(|00...0\rangle _{12...m}-|11...1\rangle
_{12...m})}^{\otimes n}.
\end{eqnarray}%
The $\left( n,m\right) $-QPC consists of $n$ sub-blocks, and each sub-block
has $m$ physical qubits. First, we define the Pauli operators, $%
X_{i,j},Y_{i,j},Z_{i,j}$ associated the $\left( i,j\right) $-th qubit, where
$i=1,\cdots ,n$ is the row (sub-block) label and $j=1,\cdots ,m$ is the
column label for the qubit. There is one logical qubit encoded in the $%
\left( n,m\right) $-QPC, with logical operators $\tilde{Z}\equiv
\prod_{i=1}^{n}Z_{i,j}$ and $\tilde{X}\equiv \prod_{j=1}^{m}X_{i,j}$, where
we may choose any $j=1,\cdots ,m$ for $\tilde{Z}$ and any $i=1,\cdots ,n$
for $\tilde{X}$ \cite{Nielsen}. The encoded states $\{|0\rangle
_{L},|1\rangle _{L},|+\rangle _{L},|-\rangle _{L}\}$ can be prepared
fault-tolerantly with with suppressed correlated errors \cite{Supp,Brooks,Hayes}. The encoded state is transmitted via an
optical fiber to the neighboring repeater station followed by error
correction and transmission to the next repeater station (Fig.~\ref%
{fig:repeater}).

Suppose that each transmitted physical qubit can reach the next QR station
with probability $\eta $, meanwhile suffering from depolarization errors. We
apply TEC \cite{Tele, KLM} to correct both photon loss and depolarization
errors. The TEC procedure consists of Bell state preparation and Bell
measurement at the encoded level (Fig.~\ref{fig:repeater} (b) ), and both
operations can be achieved fault-tolerantly without propagating errors
within each encoding block \cite{Nielsen}. The Bell measurement of two
encoded blocks (received block $R$ and local block $S$) can be achieved by
an encoded CNOT gate followed by measurement of logical operators $\tilde{X}%
^{R}$ and $\tilde{Z}^{S}$. More specifically, it consists of $nm$ pair-wise
CNOT gates between $R_{i,j}$ and $S_{i,j}$, followed by $2nm$ individual
qubit measurements. Besides erasure errors, TEC can also correct
operational errors from qubit depolarization ($\varepsilon _{d}$), imperfect
measurement ($\varepsilon _{m}$), and noisy quantum gates ($\varepsilon _{g}$%
), which can be captured by an effective error probability $\varepsilon
=\varepsilon _{d}+\frac{\varepsilon _{g}}{2}+2\varepsilon _{m}+O(\varepsilon
_{d,g,m}^{2})$ acting on single qubit for our fault-tolerant circuit designs
\cite{Supp}.

\begin{figure}[tbp]
\centering
\par
\includegraphics[width=8.5cm,angle=0]{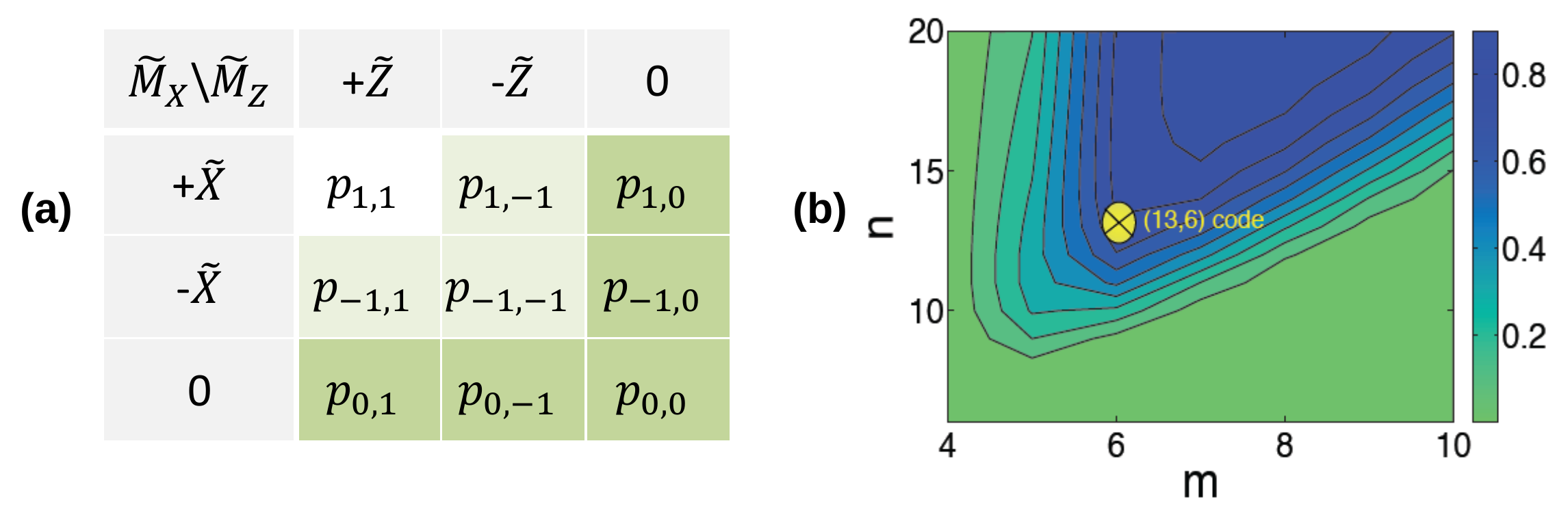}
\caption[fig:rateplot4]{(color online). (a) Distribution of possible
measurement outcomes. Measurement outcomes in the lighter area ($p_{1,-1}$, $%
p_{-1,1}$, $p_{-1,-1}$) are logical errors, and outcomes in the darker area (%
$p_{\protect\alpha ,0}$ and $p_{0,\protect\beta }$) lead to heralded
failure. (b) Contour plot for the key generation rate $R(n,m)$ (with $%
t_{0}=1 $) across a total distance $L_{tot}=10,000\mbox{km}$ with repeater
spacing $L_{0}=1.5\mbox{km}$ and $\protect\varepsilon =10^{-3}$. The
optimized choice of encoding with minimum cost [see Eq.~(\protect\ref%
{eq:Copt})] is a (13,6) code. }
\label{fig:rateplot4}
\end{figure}

In the presence of photon loss errors, each measurement may have three
possible outcomes $\left\{ +1,-1,0\right\} $. Each qubit $R_{i,j}$ in the $R$
block is measured in the X basis with outcome $X_{i,j}^{R}$ taking value $+1$
for qubit state $\left\vert +\right\rangle $, $-1$ for qubit state $%
\left\vert -\right\rangle $, and $0$ if the qubit is erased due to
transmission loss. Similarly, each qubit $S_{i,j}$ from the $S$ block is
measured in the Z basis with outcome $Z_{i,j}^{S}$ taking value $+1$ for
qubit state $\left\vert 0\right\rangle $, $-1$ for qubit $\left\vert
1\right\rangle $, and $0$ if the corresponding qubit in the $R$-block ($%
R_{i,j}$) is erased. The logical measurement outcomes depend on individual
qubit measurement outcomes
\[
{\small \tilde{M}_{X}^{R}=\text{\textrm{sign}}\left[ \sum_{i=1}^{n}\left(
\prod_{j=1}^{m}X_{i,j}^{R}\right) \right] ,~\tilde{M}_{Z}^{S}=%
\prod_{i=1}^{n}\left( \text{\textrm{sign}}\left[ \sum_{j=1}^{m}Z_{i,j}^{S}%
\right] \right) ,}
\]%
with three possible values $\left\{ +1,-1,0\right\} $. Here $\mathrm{sign}%
\left[ \sum \cdots \right] $ is associated with majority voting between $%
\left\{ \pm 1\right\} $, and $\prod \cdots $ is associated with the product
of trinary outcomes. Ideally, in the absence of errors, the outcomes should
be $\tilde{M}_{X}^{R}=\tilde{X}^{R}$ and $\tilde{M}_{Z}^{S}=\tilde{Z}^{S}$,
which determine the Pauli frame \cite{Tele,KLM} of the encoded block after
teleportation. In the presence of errors, however, the outcomes become $%
\tilde{M}_{X}^{R}=\alpha \tilde{X}^{R}$ and $\tilde{M}_{Z}^{S}=\beta \tilde{Z%
}^{S}$, with $\alpha ,\beta =+1$ for correct measurement, $-1$ for incorrect
measurement, and $0$ for heralded failure of measurement. We obtain the
probability distribution (see Fig. \ref{fig:rateplot4}a) \cite{Supp}
%\begin{equation}
$p_{\alpha ,\beta }\equiv \mathrm{Pr}\left[ \tilde{M}_{X}^{R}=\alpha \tilde{X%
}^{R},\tilde{M}_{Z}^{S}=\beta \tilde{Z}^{S}\right] $, %\end{equation}%
which can be used to evaluate the QR performance.

\paragraph*{Quantum bit error rate and success probability}

We use the probability distribution to compute the success probability and
quantum bit error rate (QBER) that characterizes the QR. Since the encoded
qubit passes through $N$ repeater stations, there are $N$ pairs of
measurement outcomes ($\tilde{M}_{X}$ and $\tilde{M}_{Z}$). The success
probability with no heralded failure of measurements is
\begin{equation}
P_{succ}=\left( p_{1,1}+p_{1,-1}+p_{-1,1}+p_{-1,-1}\right) ^{N}.
\end{equation}%
Given that all measurement outcomes have no heralded failure, there might be
an odd number of incorrect measurements of $\tilde{M}_{X}$ (or $\tilde{M}%
_{Z} $), which induces an error if the receiver decodes the information by
measuring $\tilde{X}$ (or $\tilde{Z}$) of the received block. We define the
QBER at the encoded level of the QR as the ratio of the probability of
having an odd number of incorrect measurements of $\tilde{M}_{X}$ (or $%
\tilde{M}_{Z}$) to the probability of having no heralded failure. The QBER
for $\tilde{X}$ (or $\tilde{Z}$) measurement by the receiver is
\begin{equation}
Q_{(X/Z)}=\frac{1}{2}\left[ 1-\left( \frac{p_{1,1}\pm p_{1,-1}\mp
p_{-1,1}-p_{-1,-1}}{p_{1,1}+p_{1,-1}+p_{-1,1}+p_{-1,-1}}\right) ^{N}\right] .
\end{equation}

\paragraph*{Key generation rate.}

For our QR, the raw key generation rate is $1/t_{0}$, where $t_{0}$ is the
time taken for TEC. For simplicity, we may use $t_{0}$ as a time unit in our
analysis. The raw keys can be converted to secure keys through classical
communication protocols involving error correction and privacy amplification
\cite{Qcryp}. Due to finite success probability and non-vanishing QBER, the
asymptotic secure key generation rate is given by \cite{Key1, Key2}
\begin{equation}
R=\text{max}\left[ \frac{1}{t_{0}}P_{succ}\left( 1-2h(Q)\right) ,0\right] ,
\end{equation}%
where $Q=\frac{(Q_{X}+Q_{Z})}{2}$ and $h(Q)=-Qlog_{2}(Q)-(1-Q)log_{2}(1-Q)$
is the binary entropy function. In Fig.~\ref{fig:rateplot4}, we show that $R$
can approach $1/t_{0}$ for reasonable encoding size $(n\times m)$ with an
appropriate repeater spacing $(L_{0})$, because it is possible to achieve
high $P_{succ}$ and low $Q$.
The range of $(n,m)$ that yields a high key generation rate varies with $%
L_{0}$ and the total distance of communication $L_{tot}(=N\times L_{0}$).
Hence, we need to optimize the repeater parameters, including the size of
encoding block, repeater spacing, and secure key generation rate.

For each secret bit generated by the QR, we should consider the cost of both
time and qubit resources \cite{opt}: (1) the average time to generate a
secret bit is $1/R$, and (2) the total number of memory qubits needed for
the QR scheme is $2mn\times \frac{L_{tot}}{L_{0}}$ \footnote{%
There is an overhead of ancillary qubits to enable the fault tolerant preparation of the
encoded Bell pair. This overhead depends on the fault tolerant preparation scheme as discussed in \cite{Supp}.
To fix ideas, we will use the number of memory qubits $2mn$ in our calculations.%
}. We introduce the \emph{cost function}, $C$ to be the product
of these two factors $\frac{2nm}{R}\times \frac{L_{tot}}{L_{0}}$, in units
of [qubits $\cdot t_{0}$/sbit]. Here the rate $R$ implicitly depends on the
control parameters of $\{n,m,L_{0}\}$. For given $L_{tot}$, we can achieve
the minimum cost:
\begin{equation}
C(L_{tot})\equiv \min_{n,m,L_{0}}\frac{2nm}{R}\times \frac{L_{tot}}{L_{0}},
\label{eq:Copt}
\end{equation}%
among all possible choices of $\left( n,m\right) $-QPC and repeater spacing $%
L_{0}$. We assume the following imperfections as we search for the optimal
scheme:\textbf{\ }(1) operation error with probability $\varepsilon $,
and (2) finite photon transmission with probability $\eta
=(1-p_{c})e^{-L_{0}/L_{att}}$ due to fiber attenuation ($L_{att}=20$km) and
coupling loss ($p_{c}$).

\paragraph*{Numerical search for optimized strategy.}

\begin{figure}[h]
\centering
\includegraphics[width=8.2cm]{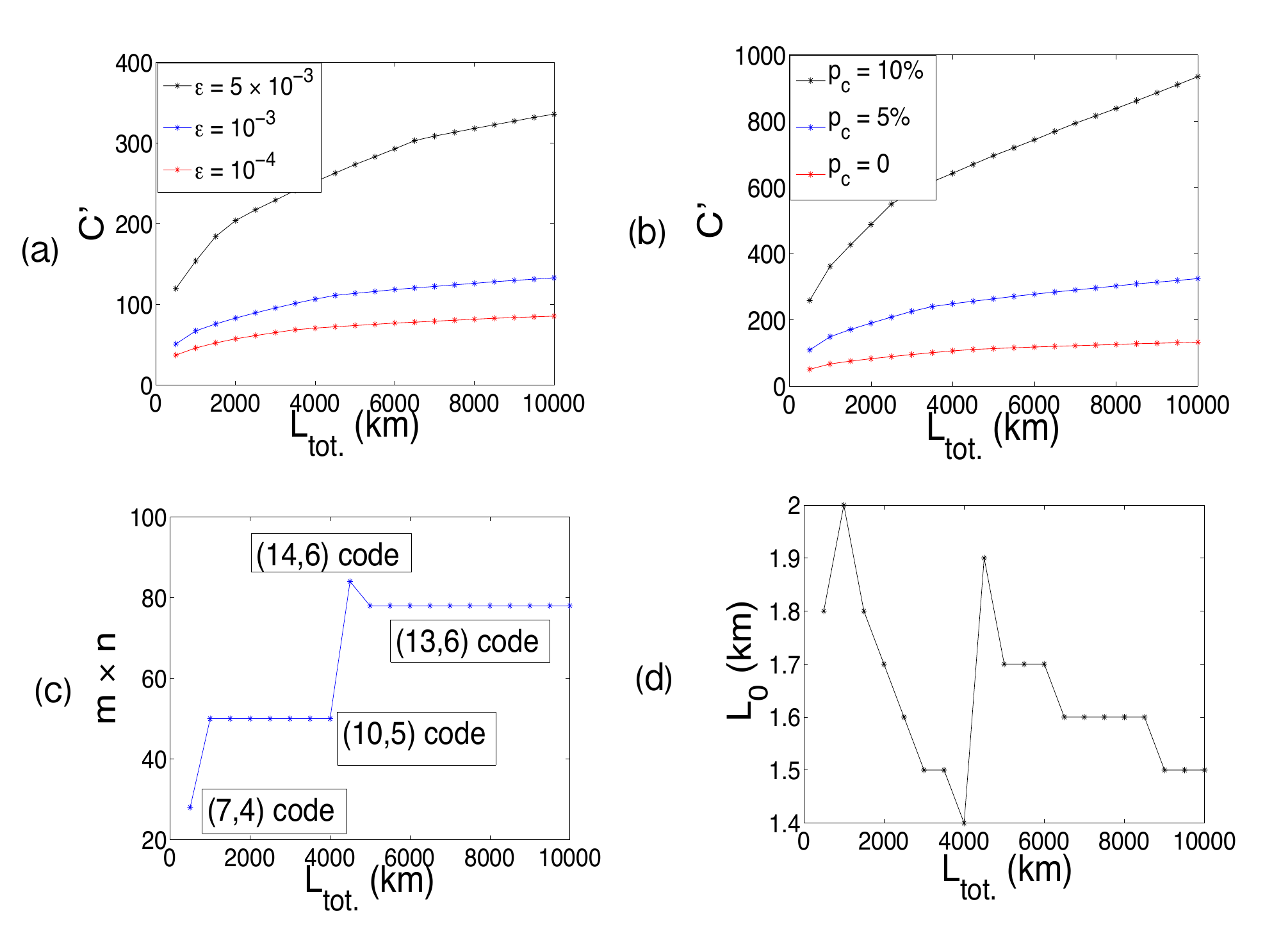}
\caption[fig:app7]{(color online). (a) Cost coefficient $C^{\prime }\left(
L_{tot}\right) $ for different operational errors $\protect\varepsilon $
with zero coupling loss $p_{c}=0$, (b) cost coefficient $C^{\prime }\left(
L_{tot}\right) $ for varying $p_{c}$ with fixed $\protect\varepsilon %
=10^{-3} $, (c,d) optimized encoded block size and repeater spacing for $%
\protect\varepsilon =10^{-3}$ and $p_{c}=0$.}
\label{fig:app7}
\end{figure}

We search for optimized choices of $\left\{ n,m,L_{0}\right\} $ for
different values of $L_{tot}$ with fixed imperfection parameters of $\left\{
\varepsilon ,p_{c}\right\} $. We run a numerical search for $L_{0}$ and for different number of qubits
to obtain $C\left( L_{tot}\right) $, which should
increase at least linearly with $L_{tot}$. In Fig.~\ref{fig:app7}, we show
the variation of \emph{cost coefficient} $(C^{\prime }=C/L_{tot})$ with $%
L_{tot}$, to illustrate the additional overhead associated with $L_{tot}$.
The cost coefficient can be interpreted as the resource overhead (qubits$%
\times t_{0}$) for the creation of one secret bit over 1km (with target distance
$L_{tot}$).

For imperfection parameters of $\epsilon =10^{-3}$ and $p_{c}=0$, the
algorithm picks only four different codes up to $L_{tot}=10,000\mbox{km}$.
When the code chosen by the algorithm changes (for example at $4500\mbox{km}$
in Fig.~\ref{fig:app7}), the product of $L_{0}$ and $R$ also jumps to an
appropriate higher value, so that the cost coefficient varies continuously
with $L_{tot}$. In the presence of coupling loss $p_{c}<10\%$, the optimized
values of $L_{0} $ is within the range $1.4\sim 2\mbox{km}$ (Fig.~\ref%
{fig:app7}) with total loss errors up to $20\%$; $R\cdot t_{0}$ is high ($%
0.6-0.85$) because of the favorable QBER associated with the chosen codes.

The optimized cost coefficient for different operational error probabilities
is shown in Fig.~\ref{fig:app7}. When $\varepsilon $ decreases below $%
10^{-3} $, the cost coefficient is dominated by photon loss errors rather
than operational errors, and does not decrease by a significant amount as $%
\varepsilon $ decreases further. In a realistic scenario, photons are lost
due to finite coupling losses besides fiber attenuation. In Fig.~\ref%
{fig:app7}, we show that the QR scheme can tolerate coupling losses up to $%
10\%$ for a reasonable overhead in the number of qubits. Numerical
calculations indicate that the cost coefficient increases by $%
O(poly(log(L_{tot})))$ \cite{Supp}. Table \ref{tab:resource} provide an
estimate of the resource requirements of our code under different scenarios.
%
%An estimate of the resource requirements
%of our code under different scenarios is shown in Table \ref{tab:resource}.

\paragraph*{Experimental considerations.}

To implement our QR scheme, it is crucial to fulfill the following two
experimental requirements: (1) The coupling loss should be sufficient low ($%
p_{c}\lesssim 10\%$), because if the transmission
probability $\eta <50\%$, then the probability that the receiver decodes the
logical qubit will be exponentially small \footnote{%
Our repeater scheme is effectively a sequence of quantum erasure channels
with forward-only communication between neighboring repeater stations, which
has capacity $\max \left[ 0,2\eta-1\right] $ \cite{Bennett97}. If $\eta <1/2
$, the channel capacity vanishes, which implies that the probability of
faithful transmission between neighboring repeater stations cannot approach
unity, and consequently the probability of faithful transmission over many
repeater stations will decrease exponentially with the number of repeater
stations.}. (2) Quantum repeater station should have hundreds of qubits with high
fidelity operations. For ion trap systems, single qubit gate error
probability of $2\times 10^{-5}$ \cite{qgate4}, two-qubit gate error
probability of $0.007$ \cite{qgate3}, and measurement error probability of $%
10^{-4}$ \cite{qgate5,qgate31} have been demonstrated.
There are also promising developments in micro-fabricated ion traps for
coherent control of hundreds of ion qubits \cite{Monroe13}.

In addition to these two requirements, efficient downconversion to telecom
wavelength (using similar techniques as described in \cite{spd}, where
conversion efficiency of up to $86\%$ was reported) can be used to
minimize fiber attenuation. The collection efficiency of the photon from the
ion (enhanced by adequate cavity QED effect \cite{coll,fiber}), wavelength
conversion efficiency, and coupling of the resulting photons into the
propagating media (fiber) should all be maximized to $90\%$ levels, which
remains an experimental challenge.

The techniques of time and wavelength-division-multiplexing will enable us
to transmit multiple photons through a single optical fiber, increasing the
communication rate by as much as four orders of magnitude ($100$
wavelengths, with $100$ ions transmitting in sequence). The operation of TEC
can be achieved with cavity QED systems \cite{Duan04b,Supp}. The performance
of the QR scheme introduced here depends crucially on the range of input
parameters $(\varepsilon ,p_{c},t_{0})$. The key generation rate $R$ depends
on the TEC time of $t_{0}$. Since it is possible to have sub-nanosecond
quantum gates \cite{qgate1,qgate2} with trapped ions, the TEC time will be
mostly limited by the relatively slow measurement ($10\sim 100\mu s$) \cite%
{detection} due to finite photon scattering rate and collection efficiency,
which can be significantly improved by enhancing the ion-cavity coupling
strength. For instance, if the TEC time is improved to $t_{0}=1\ \mu s$, a
secure key generation rate over $0.5$ MHz can be achieved over $10,000$ km
with the $(41,8)$ code for $\varepsilon =10^{-3}$, $p_{c}=10\%$ and $L_{0}=1.2$km.

Besides trapped ions, neutral atoms in cavities \cite{Rempe, atomluk}, NV centers \cite{NV1, NV2}, quantum dots \cite{QD1,QD2}, and Rydberg atoms \cite{Ryd1, Ryd2} are also promising systems for quantum repeater implementations. Furthermore, with the advance of coherent
conversion between optical and microwave photons \cite{micro},
superconducting qubits may become an attractive candidate to realize our
scheme as they can achieve both ultrafast quantum gates and high coupling
efficiency.
\begin{table}[tbp]
\centering%
\begin{tabular}{c|c|cccc|cccc}
\hline\hline
$p_{c}$ & $\varepsilon $ & \multicolumn{4}{|c|}{$L_{tot}=1000$\mbox{km}} &
\multicolumn{4}{|c}{$L_{tot}=10000$\mbox{km}} \\
&  & $m$ & $n$ & $L_{0}$ (km) & $R\cdot t_{0}$ & $m$ & $n$ & $L_{0}$ (km) & $%
R\cdot t_{0}$ \\ \hline
$0\%$ & $10^{-4}$ & $4$ & $7$ & $1.7$ & $0.72$ & $5$ & $9$ & $1.3$ & $0.80$
\\
$0\%$ & $10^{-3}$ & $5$ & $10$ & $2.0$ & $0.74$ & $6$ & $13$ & $1.5$ & $0.78$
\\
$10\%$ & $10^{-4}$ & $6$ & $21$ & $1.6$ & $0.60$ & $7$ & $28$ & $1.0$ & $%
0.57 $ \\
$10\%$ & $10^{-3}$ & $7$ & $31$ & $1.8$ & $0.67$ & $8$ & $41$ & $1.2$ & $%
0.59 $ \\ \hline\hline
\end{tabular}%
\caption{Optimized resource requirements for our fault-tolerant QR scheme
with $(n,m)$-QPC encoding for different coupling losses $p_{c}$ and
operational error $\protect\varepsilon $.}
\label{tab:resource}
\end{table}

\paragraph*{Summary and Outlook.}

We have presented a new QR scheme belonging to the third class of QRs, which
considers both fault tolerance and small encoding blocks for ultrafast
quantum communication over long distances. In comparison with the first and
second classes of QR schemes, our QR scheme uses TEC within each QR station
to correct both photon loss and operation errors. In particular, our QR
scheme can tolerate finite coupling loss ($p_{c}\lesssim 10\%$) and achieve
fault-tolerant operation with approximately hundreds of qubits per repeater
station. This enables improved key generation rate that is limited only by
local gate operations. Our scheme requires smaller QR spacing compared to
the previous classes of QRs and consequently the number of QR stations is
higher by roughly an order of magnitude. But it is important to note that
the key generation rate increases by more than three orders of magnitude, by
eliminating the communication time between the repeater stations. In
addition, we have introduced a cost function to optimize the control
parameters of our QR scheme, which can potentially be used as a criterion to
compare all three classes of QRs as well as to search for more efficient
quantum error correcting codes for quantum communication.

\paragraph*{Acknowledgements}
This work was supported by the DARPA (Quiness program), NSF, CUA, NBRPC (973 program),
Packard Foundation, Alfred P. Sloan Foundation. We thank Austin Fowler, Steven Girvin, Archana
Kamal, Chris Monroe, Bill Munro, David Poulin and Hong Tang for discussions.

%%%%%%%%%%%%%%%%%%%%%%%%%%%%%%%%%%%%%%%%%%%%%%%%%%%%%%%%%%%%%%%
%%%%%%%%%%   Supplemental Material     %%%%%%%%%%%%%%%%%%%%%%%%
%%%%%%%%%%%%%%%%%%%%%%%%%%%%%%%%%%%%%%%%%%%%%%%%%%%%%%%%%%%%%%%
%\newpage
%
%\title{Supplemental Material: \\
%Ultrafast and Fault-Tolerant Quantum Communication across Long Distances}
%\author{Sreraman Muralidharan$^{1}$}
%\author{Jungsang Kim$^{2}$}
%\author{Norbert L\"utkenhaus$^{3}$}
%\author{Mikhail D. Lukin$^{4}$}
%\author{Liang Jiang$^{5}$}
%\affiliation{$^1$Department of Electrical Engineering, Yale University, New Haven, CT
%06511 USA}
%\affiliation{$^2$Department of Electrical and Computer Engineering, Duke University,
%Durham, NC 27708 USA}
%\affiliation{$^3$Institute of Quantum computing, University of Waterloo, N2L 3G1
%Waterloo, Canada}
%\affiliation{$^4$Department of Physics, Harvard University, Cambridge, MA 02138, USA}
%\affiliation{$^5$Department of Applied Physics, Yale University, New Haven, CT 06511 USA}
%%\date{\today }
%%\pacs{03.67.Dd, 03.67.Hk, 03.67.Pp.}

\section{Supplemental Material}
In the Supplemental Material, we first provide an overview of all three
classes of Quantum repeaters (QRs). Then we present key procedures of
fault-tolerant preparation of the encoded quantum states,
teleportation-based error correction (TEC) and its implementation in
cavity-QED systems. After that we give an in-depth analysis of various
errors and calculate the probability distribution of measurement outcomes at
each repeater station. Finally, we provide the optimization algorithm and
discuss the scaling of the cost function with respect to the long distance
of communication.

\section{Introduction}

The first two classes of QRs require generation of heralded EPR pairs
between neighboring repeater stations, and entanglement purification or
quantum error correction steps to generate an EPR pair of high fidelity
between distant repeater stations. If a photon is lost in the procedure, the
heralded outcome will be a failure and the procedure will be repeated until
it succeeds. Hence, apart from a constant time overhead, photon loss events
do not have a major role to play in the success or the failure rates of the
protocol. However, the heralded outcome requires two-way classical
communication, which limits the key generation rate of the first two classes
of QRs. In our new scheme for QRs, entanglement purification steps (in the
first class of QRs) and the heralded entanglement generation steps (in the
second class of QRs) are replaced by quantum error correction at individual
repeater stations, eliminating the need to establish entangled links between
any two repeater stations. Such a procedure makes use of one way classical
communication which can be very fast and only limited by the speed of local
operations. On the other hand, photon losses during the transmission of the
encoded state may lead to failure in the secret key generation. Therefore,
it is important that the error correction procedure at individual repeater
stations can correct both loss and operational errors.

Keeping this in mind, we choose the Calderbank-Shor-Steane (CSS) encoding
because they have properties of fault-tolerant state preparation and gate
implementation. For instance, an encoded CNOT gate between the codewords can
be achieved by simply applying transversal CNOT gates between the physical
qubits. As we will see later, this is essential to perform a fault-tolerant
quantum error correction at individual repeater stations.

\section{Fault-tolerant preparation of the encoded quantum states}
In this section, we will provide key procedures to prepare encoded states of
quantum parity code (QPC) within each repeater station. We assume that
within each repeater station there are long range interconnects for state
preparation, making the physical location of the qubits irrelevant (e.g.,
this can be achieved in an anharmonic linear ion trap \cite{Lin09}). In
principle, the standard procedure for fault-tolerant preparation of CSS
codes \cite{NC00} can be applied to our QPC encoding, because QPC is a
special class of CSS code. For completeness, we will provide the procedure
of fault-tolerant preparation of QPC, because the logical operators and
stabilizer of QPC have special structures which enables efficient state
preparation.

We define the $\left( n,m\right) $-QPC using the stabilizer formalism \cite%
{Stabilizer}. We use the Pauli operators $X_{i,j},Y_{i,j},Z_{i,j}$ for the $%
\left( i,j\right) $-th qubit, where $i=1,\cdots ,n$ is the row (sub-block)
label and $j=1,\cdots ,m$ is the column label for the qubit. The stabilizer
operators for the $\left( n,m\right) $-QPC are%
\begin{equation*}
S_{i,j}\equiv Z_{i,j}Z_{i,j+1}
\end{equation*}%
with $i=1,\cdots ,n$ and $j=1,\cdots ,m-1$, and%
\begin{equation*}
S_{i,0}\equiv \prod_{j=1}^{m}X_{i,j}X_{i+1,j}
\end{equation*}%
with $i=1,\cdots ,n-1$. Given the above $nm-1$ independent stabilizer
operators, there is one logical qubit encoded in the $\left( n,m\right) $%
-QPC, with logical operators%
\begin{equation*}
\tilde{Z}\equiv \prod_{i=1}^{n}Z_{i,1}=\prod_{i=1}^{n}Z_{i,2}=\cdots
=\prod_{i=1}^{n}Z_{i,m}
\end{equation*}%
\begin{equation*}
\tilde{X}\equiv \prod_{j=1}^{m}X_{1,j}=\prod_{j=1}^{m}X_{2,j}=\cdots
=\prod_{j=1}^{m}X_{n,j}.
\end{equation*}%
The distance of the code is given by $d=min(n,m)$. In the following, we will
focus on fault-tolerant preparation of three encoded quantum states --- $%
|0\rangle _{L}$, $|+\rangle _{L}$, and $\frac{1}{\sqrt{2}}(|00\rangle
_{L}+|11\rangle _{L})$ --- which are needed for our new scheme of QRs.

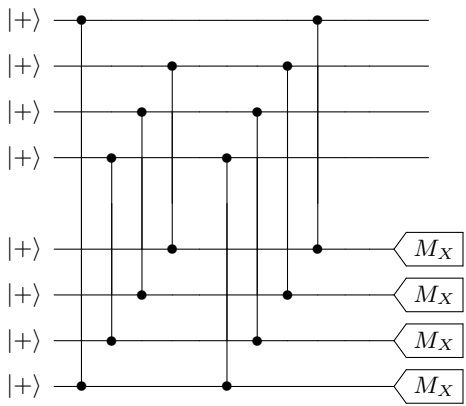
\begin{figure}[h]
\mbox{
\Qcircuit @C=1em @R=0.5em @!R {
\lstick{\ket{+}}& \ctrl{7} & \qw & \qw & \qw & \qw & \qw & \qw & \qw & \ctrl{4} & \qw & \qw & \qw\\
\lstick{\ket{+}}& \qw & \qw &  \qw & \ctrl{3} & \qw & \qw & \qw & \ctrl{4} & \qw & \qw & \qw & \qw\\
\lstick{\ket{+}}& \qw & \qw  & \ctrl{3} & \qw & \qw & \qw & \ctrl{4} & \qw & \qw & \qw & \qw & \qw\\
\lstick{\ket{+}}& \qw & \ctrl{3} & \qw & \qw & \qw & \ctrl{4} & \qw & \qw & \qw & \qw & \qw & \qw\\
 & & &&&&&& &&&\\
\lstick{\ket{+}}& \qw & \qw & \qw & \ctrl{-3} & \qw & \qw & \qw & \qw & \ctrl{-4} & \qw & \qw & \measuretab{M_{X}}\\
\lstick{\ket{+}}& \qw & \qw & \ctrl{-3} & \qw & \qw & \qw & \qw & \ctrl{-4} & \qw & \qw & \qw& \measuretab{M_{X}}\\
\lstick{\ket{+}}& \qw & \ctrl{-3} & \qw & \qw & \qw & \qw & \ctrl{-4} & \qw & \qw & \qw & \qw& \measuretab{M_{X}}\\
\lstick{\ket{+}}& \ctrl{-7} & \qw & \qw & \qw & \qw & \ctrl{-4} & \qw & \qw & \qw & \qw & \qw& \measuretab{M_{X}}
}}
\caption[Fig:syn]{Fault-tolerant preparation of a GHZ state, following the
scheme by Brooks and Preskill \protect\cite{Brooks} . The syndrome
measurements in the circuit are repeated $r^{\prime }$ times.}
\label{Fig:syn}
\end{figure}

First, we can fault-tolerantly prepare the encoded state $|+\rangle _{L}=%
\frac{1}{2^{n/2}}{(|00...0\rangle _{12...m}+|11...1\rangle _{12...m})}%
^{\otimes n}$, which is simply a tensor product of $n$ copies of $m$-qubit
GHZ states (also called cat states). There are many approaches to prepare
the $m$-qubit GHZ states fault-tolerantly. To fix ideas, we outline the
preparation-verification procedure provided by Brooks and Preskill \cite%
{Brooks}: (1) prepare the product state $|+\rangle ^{\otimes m}$; (2)
measure the $m$ $ZZ$ stabilizer operators using $m$ ancilla qubits as
illustrated in Fig. \ref{Fig:syn} \footnote{%
Conditioned on an even number $-1$ outcomes, the measurement syndrome is
used to estimate the "Pauli frame" of the computation.}; (3) repeat step $%
r^{\prime }$ times to suppress measurement errors; (4) determine the
syndrome of the prepared GHZ state by picking the syndrome that occurs most
frequently (or performing a perfect matching algorithm) based on the
space-time history of the syndrome measurement.
The syndrome associated with
$X$ errors need not be corrected, because we can track their propagation as
the computation proceeds, by updating the \textquotedblleft Pauli
frames\textquotedblright\ \cite{Knill05} of the individual physical qubits.
(A detailed error analysis of the GHZ state preparation is presented in \cite%
{Brooks}.) Following the above procedure, we can prepare $n$ independent
copies of $m$-qubit GHZ states, and obtain the fault-tolerant preparation of
the logical state $|+\rangle _{L}$.

\begin{figure}[h]
\mbox{
\Qcircuit @C=1em @R=0.5em @!R {
& & &&&\ctrl{4} & \qw & \qw & \qw & \qw & \qw & \qw \\
& &&&&\qw & \ctrl{4} & \qw & \qw & \qw & \qw & \qw \\
& &&&&\qw & \qw & \ctrl{4} & \qw & \qw & \qw & \qw \\
& &&&&\qw & \qw & \qw & \ctrl{4} & \qw & \qw & \qw  \\
&&&&&&&&&&&&&&\\
&&&&&\ctrl{-4} & \qw & \qw & \qw & \qw & \qw & \qw &\measuretab{M_{X}} \\
\mbox{GHZ state}& &&&& \qw & \ctrl{-4}& \qw & \qw & \qw & \qw & \qw & \measuretab{M_{X}}\\
& &&&&\qw & \qw & \ctrl{-4} & \qw & \qw & \qw & \qw &\measuretab{M_{X}}\\
&&& &&\qw & \qw & \qw & \ctrl{-4} & \qw & \qw & \qw& \measuretab{M_{X}}
  \gategroup{9}{9}{6}{5}{.7em}{\{}}
}
\caption[Fig:syn2]{The measurement of the stabilizer $S_{i,0}$ using a GHZ
state. One needs a $2m$ qubit GHZ state to measure the stabilizers which are
associated with two-consecutive rows of the QPC. This measurement is repeated $r''$ times.}
\label{Fig:syn2}
\end{figure}
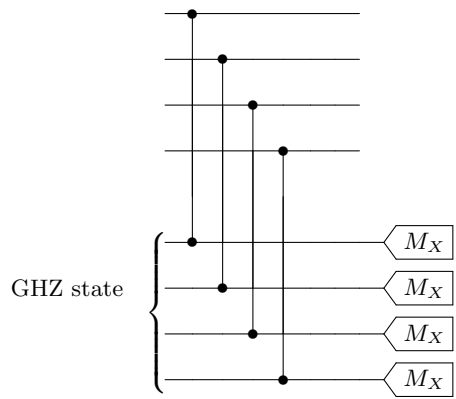

We can also fault-tolerantly prepare the encoded state $|0\rangle _{L}$.
Different from $|+\rangle _{L}$ discussed earlier, $|0\rangle _{L}$ cannot
be decomposed as a tensor product of some simple GHZ states. Hence, we
follow the standard procedure of fault-tolerant preparation for CSS codes:
(1) prepare the produce state $|0\rangle ^{\otimes nm}$ to ensure $S_{i,j}=1$
and $\tilde{Z}=1$; (2) repeatedly measure the stabilizer operators $S_{i,0}$
using $2m$-qubit GHZ states \ref{Fig:syn2}, which can be fault-tolerantly
prepared as discussed earlier; (3) repeat step $r''$ times to suppress
measurement errors; (4) determine the syndrome associated with stabilizer
operators $S_{i,0}$ based on the space-time history of the syndrome
measurement. Note that gate errors during the syndrome extraction will not
cause correlated errors in the encoding block, as each quantum gate can
affect at most one physical qubit from the encoding block. The syndrome need
not be corrected, as we can track their progagation by updating the Pauli
frames. Therefore, we can prepare the logical state $|0\rangle _{L}$
fault-tolerantly. Following the analysis of Brooks and Preskill \cite{Brooks}, upper bounds on the errors in the preparation of a GHZ state $P_{\mbox{err}}(\mbox{GHZ})$
and the probability that atleast one of the stabilizers is decoded wrongly $P_{\mbox{err}}(\mbox{Stabilizer})$ can be determined for different values of $r'$ and $r''$ as shown in Table \ref{tab:stabilizer}.

\begin{table}[t]
\begin{center}
{\normalsize
\begin{tabular}{|l|l|l|l|l|l}
\hline
Code & {$r'$} & $r''$ & $P_{\mbox{err}}(\mbox{GHZ})$ & $P_{\mbox{err}}(\mbox{Stabilizer})$ \\ \hline
$(13,6)$ & {$4$} & $11$ & $10^{-10}$ & $2.9\times 10^{-4}$  \\
$(13,6)$ & {$4$} & $21$ & $3\times 10^{-10}$ & $1.9\times 10^{-7}$ \\
$(13,6)$ & {$4$} & $31$ & $8.6\times 10^{-10}$ & $1.4\times 10^{-10}$\\ \hline
$(7,4)$ & {$4$} & $11$ & $4\times 10^{-12}$ & $1.2\times 10^{-5}$ \\
$(7,4)$ & {$4$} & $21$ & $1.5\times 10^{-11}$ & $1.1\times 10^{-9}$ \\
$(7,4)$ & {$4$} & $31$ & $3.3\times 10^{-11}$ & $10^{-13}$ \\
\hline
\end{tabular}%
}
\end{center}
\caption[tab:stabilizer]{An estimate of the upper bounds of $P_{\mbox{err}}(\mbox{GHZ})$ and $P_{\mbox{err}}(\mbox{Stabilizer})$ for different codes for different number of rounds of syndrome measurements with gate error $\epsilon_g = 10^{-3}$ and measurement error $\epsilon_m=10^{-4}$. }
\label{tab:stabilizer}
\end{table}
Finally, we can fault-tolerantly prepare the encoded Bell state $|\Phi
^{+}\rangle _{L}=\frac{1}{\sqrt{2}}(|00\rangle _{L}+|11\rangle _{L})$ by
applying encoded CNOT gates (i.e., transversal CNOT gates between the $%
k^{th} $ qubit of the first block and the $k^{th}$ qubit of the second block
for all $k$) between two encoding blocks $|+\rangle _{L}$ and $|0\rangle
_{L} $.
\color{black}
\section{Teleportation-based error correction}

\begin{figure}[b]
{\normalsize \centering
\includegraphics[height=1.8in,width=3.5in,angle=0]{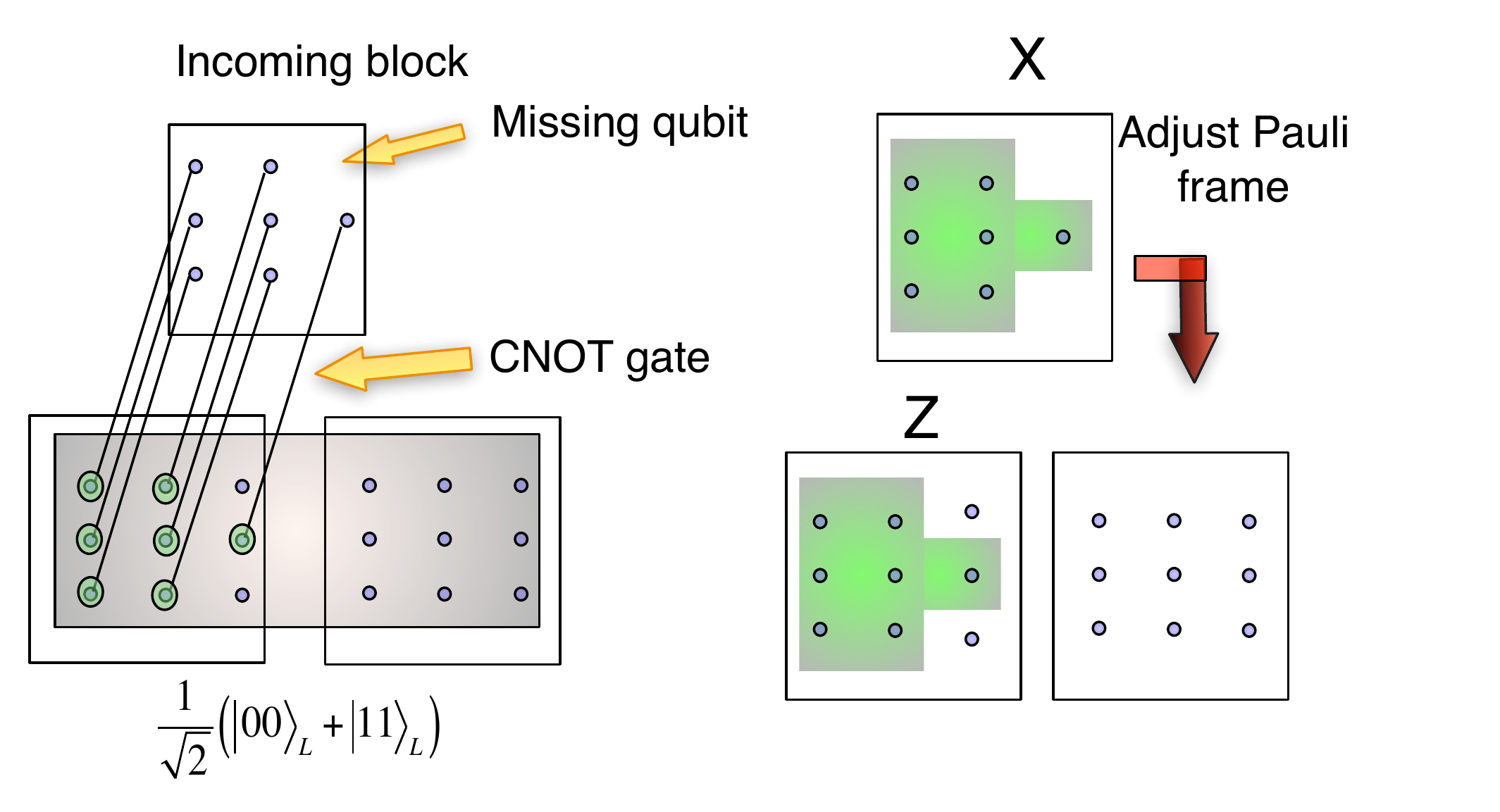} }
\caption[fig:appendix0]{(color online). An illustration of the TEC scheme
using $(3,3)$-QPC to correct the loss of two photons. The photons at
positions $(1,3)$ and $(3,3)$ are missing in the first block and
consequently a CNOT gate is not applied between the first block and the
second block at those positions. Subsequently, an encoded $X$ and $Z$
-measurements are carried out and the outcomes of the measurement are used
to adjust the Pauli frame either at the same repeater station or transmitted
to the receiver station and the Pauli frame is adjusted. The grey shading
represents entanglement before the CNOT gate and the green shading
represents measurement of encoded X and Z operators.}
\label{fig:appendix0}
\end{figure}

We now consider teleportation-based error correction (TEC) at each repeater
station. As illustrated in Fig.~1(b) of the main text, the TEC protocol
consists of preparation of encoded Bell state and Bell measurement at the
encoded level. As discussed in the previous section, we can fault-tolerantly
prepare the encoded Bell state $|\Phi ^{+}\rangle _{L}$. The Bell
measurement at the encoded level can be achieved by fault-tolerant
transversal CNOT gates followed by measurement of logical $X$ and $Z$
operators.

Consider the simple case with only photon loss errors. When an encoded block
of photons reaches a repeater station, missing photons are detected through
a quantum non-destructive measurement and the remaining photons are error
corrected by the TEC protocol. As illustrated in Fig.~\ref{fig:appendix0}, a
$(3,3)$-QPC is used to correct loss of one photon in the absence of
operational errors. In order to have successful recovery of quantum
information encoded in the $\left( n,m\right) $-QPC, both of the following
two conditions should be satisfied:

\begin{enumerate}
\item At least one qubit must arrive for each sub-block;

\item At least one sub-block must arrive with no loss.
\end{enumerate}

In a realistic scenario, there are also operational errors from imperfect
memory, measurement, and quantum gates. The TEC protocol can protect the
qubits from operational errors as well as photon loss errors, by the
following procedure of measuring the logical $X$ and $Z$ operators.

For logical $X$ measurement, we uses the definition of logical operator $%
\tilde{X}\equiv \prod_{j=1}^{m}X_{i,j}$ for $i=1,\cdots ,n$. Ideally, one
complete sub-block is sufficent for $X$ measurement. However, in the
presence of photon loss and operational errors, we need to perform \emph{%
majority voting} among all outcomes from complete sub-blocks. For example,
in Table.~\ref{tab:X-Measurement} with $\left( n,m\right) =\left( 5,4\right)
$ encoding, we rearrange the encoding blocks such that the first $n^{\prime
}=2$ sub-blocks (rows) contain missing qubits, while the remaining $%
n-n^{\prime }$ sub-blocks are complete sub-blocks. All the qubits are
measured in the $X$ basis. The $i$-th complete sub-block can infer the $%
\tilde{X}$ operator by computing $X_{i}^{\ast }\equiv \prod_{j=1}^{m}X_{i,j}$%
. Finally, we use majority voting among complete sub-blocks $\left\{
X_{i}^{\ast }\right\} _{i=n^{\prime }+1,n^{\prime }+2,\cdots ,n}$ to obtain
the true value of $\tilde{X}$.

\begin{table}[t]
\begin{center}
{\normalsize
\begin{tabular}{ccccc|c}
$i\backslash j$ & \multicolumn{1}{|c}{$1$} & $2$ & $3$ & $4$ & $X_{i}^{\ast
}\equiv \prod_{j=1}^{m}X_{i,j}$ \\ \hline
$1$ & \multicolumn{1}{|c}{$X$} & $0$ & $0$ & $0$ & $0$ \\
$2$ & \multicolumn{1}{|c}{$X$} & $X$ & $0$ & $0$ & $0$ \\
$3$ & \multicolumn{1}{|c}{$X$} & $X$ & $X$ & $X$ & $X_{3}^{\ast } = \pm 1$
\\
$4$ & \multicolumn{1}{|c}{$X$} & $X$ & $X$ & $X$ & $X_{4}^{\ast }= \pm 1$ \\
$5$ & \multicolumn{1}{|c}{$X$} & $X$ & $X$ & $X$ & $X_{5}^{\ast }= \pm 1$ \\
\hline
&  &  &  &  & $\tilde{M}^{X}_{R}=$Majority $\left\{ X_{i}^{\ast }\right\}
_{i=3,4,5}$ \\
&  &  &  &  & $= \text{sign}\left[\sum_{i=1}^{n} X_{i}^{\ast}\right]$%
\end{tabular}%
}
\end{center}
\caption[tab:X-Measurement]{For $(5,4)$-QPC, the measurement strategy of $%
\tilde{X}$ based on majority voting among $\left\{ X_{i}^{\ast }\right\}
_{i=3,\cdots ,5}$.}
\label{tab:X-Measurement}
\end{table}

For logical $Z$ measurement, we may infer the encoded logical $\tilde{Z}$
operator by calculating $\prod_{i=1}^{n}Z_{i}^{\ast }$, where $Z_{i}^{\ast }$
is obtained by majority voting from the $i$-th sub-block. For example,
Table.~\ref{tab:Z-Measurement} illustrates the computation of the value for
logical $Z$ operator in the presence of loss errors.

\begin{table}[t]
\begin{center}
{\normalsize
\begin{tabular}{ccccc|c}
$i\backslash j$ & \multicolumn{1}{|c}{$1$} & $2$ & $3$ & $4$ & $Z_{i}^{\ast
}=\text{sign}\left[\sum_{j=1}^{m}Z_{i,j}\right]$ \\ \hline
$1$ & \multicolumn{1}{|c}{$Z$} & $0$ & $0$ & $0$ & $Z_{1}^{\ast }= \pm 1$ \\
$2$ & \multicolumn{1}{|c}{$Z$} & $Z$ & $0$ & $0$ & $Z_{2}^{\ast }= \pm 1$ \\
$3$ & \multicolumn{1}{|c}{$Z$} & $Z$ & $Z$ & $Z$ & $Z_{3}^{\ast }=\pm 1$ \\
$4$ & \multicolumn{1}{|c}{$Z$} & $Z$ & $Z$ & $Z$ & $Z_{4}^{\ast }=\pm 1$ \\
$5$ & \multicolumn{1}{|c}{$Z$} & $Z$ & $Z$ & $Z$ & $Z_{5}^{\ast }=\pm 1$ \\
\hline
&  &  &  &  & $\tilde{M}^{S}_{Z}=\prod_{i=1}^{n}Z_{i}^{\ast }$%
\end{tabular}
}
\end{center}
\caption[tab:Z-Measurement]{For $(5,4)$-QPC, measurement strategy of $\tilde{%
Z}$ based on $Z_{i}^{\ast }$ with $i=1,...,n$.}
\label{tab:Z-Measurement}
\end{table}

\begin{figure}[h]
\mbox{
\Qcircuit @C=1.5em @R=1.5em @!R {
\mbox{Incoming photon $R_{i.j}$} &&&&& \qw & \qw & \ctrl{1}& \qw & \measuretab{M_X}  \\
\mbox{Local matter qubit $S_{i.j}$} &&&&& \qw & \targ &  \targ & \qw &  \measuretab{M_Z}\\
\mbox{Outgoing photon $T_{i.j}$} &&&&& \qw & \ctrl{-1} &  \qw & \qw & \qw & \\ \\
}}
\caption[fig:tecimp]{The TEC quantum circuit at the level of physical
qubits. }
\label{fig:tecimp}
\end{figure}
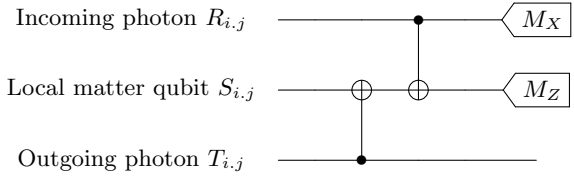

With the above procedure of measuring the logical $X$ and $Z$ operators, we
can perform the TEC fault-tolerantly. The TEC circuit at the encoded level
(Fig.1(b) in the Letter) is very similar to the TEC circuit at the physical
level (Fig.~\ref{fig:tecimp}), consisting of Bell state preparation and Bell
measurement. However, the determination of the Pauli frame is not based on
the Bell measurement at physical level, but depending on the Bell
measurement outcomes at encoded level. As shown in Fig.~\ref{fig:tecimp}, we
need to perform quantum gates that couple the incoming photon $R_{i,j}$,
local qubits $S_{i,j}$, and outgoing photon $T_{i,j}$. After that, we
measure the incoming photon $R_{i,j}$ in $X$ basis and the local atom $%
S_{i,j}$ in $Z$ basis.

Cavity QED systems can implement the TEC protocol. The key is to perform the
CNOT gate, which can be decomposed into two Hadamard gates and a CPHASE
gate, $CNOT_{a,b}=H_{b}\cdot CPHASE_{a,b}\cdot H_{b}$, with an efficiently
implementation using cavity QED systems proposed by Duan and Kimble~\cite%
{Duan04b}. For example, with polarization encoding $\{H,V\}$ for the photon,
a CHPASE gate can be achieved through an optical setup shown in Fig.~\ref%
{fig:cavity31}. Using this implementation for a CPHASE gate, the TEC circuit
can be effectively implemented for an atom inside a cavity as shown in Fig.~%
\ref{fig:cavity}.

\begin{figure}[t]
\centering
\includegraphics[width=2in]{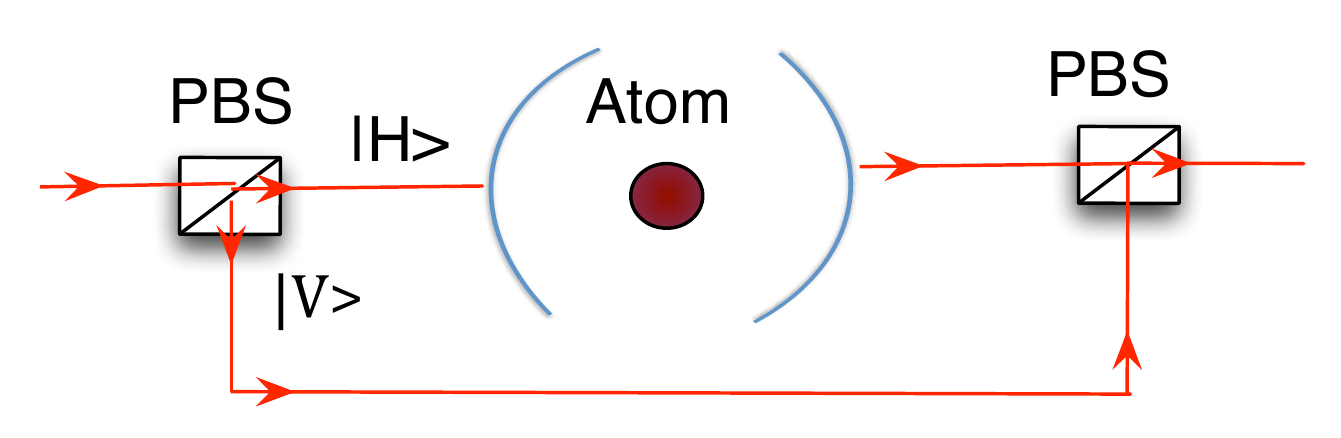}
\caption[fig:cavity31]{(color online). Implementation of a CPHASE gate
between a photon and an atom. The polarizing beam splitter (PBS) splits the
path of the input photon depending on its polarization and the atom
interacts with only photons with a certain polarization.}
\label{fig:cavity31}
\end{figure}

\section{Error Model \& Probability Distributions}

We consider error model with both photon loss and operational errors (due to
imperfect gates and measurement). Since the encoding blocks ($R$ and $S$)
are prepared fault-tolerantly and independetly, the qubits from these blocks
have independent errors before we perform the Bell measurement. Before the
application of the CNOT gate, the combined state of $R$ and $S$ can be
written as $\rho _{RS}=\rho _{R}\otimes \rho _{S}$. In the absence of CNOT
gate errors, the application of a CNOT gate operation (denoted by $U$) on
the state $\rho _{RS}$ is given by $U\rho _{RS}{U^{\dagger }}$. In the
presence of gate errors $\epsilon _{g}$, the action of a noisy CNOT gate can
be denoted with the super-operator
\begin{eqnarray}
\mathcal{E}_{RS}\left( \rho _{RS}\right)  &=&(1-\epsilon _{g})U\rho _{RS}{%
U^{\dagger }}+  \notag \\
&&\frac{\epsilon _{g}}{16}\sum_{k^{\prime }=0}^{3}\sum_{k=0}^{3}\sigma
_{k^{\prime }}^{(R)}\sigma _{k}^{(S)}\rho \sigma _{k}^{(S)}\sigma
_{k^{\prime }}^{(R)}.
\end{eqnarray}%
where $\left\{ \sigma _{k}\right\} _{k=0,\cdots ,3}=\left\{ I,X,Y,Z\right\} $
are Pauli matrices including identity$.$

\begin{figure}[t]
\centering
\includegraphics[width=3in]{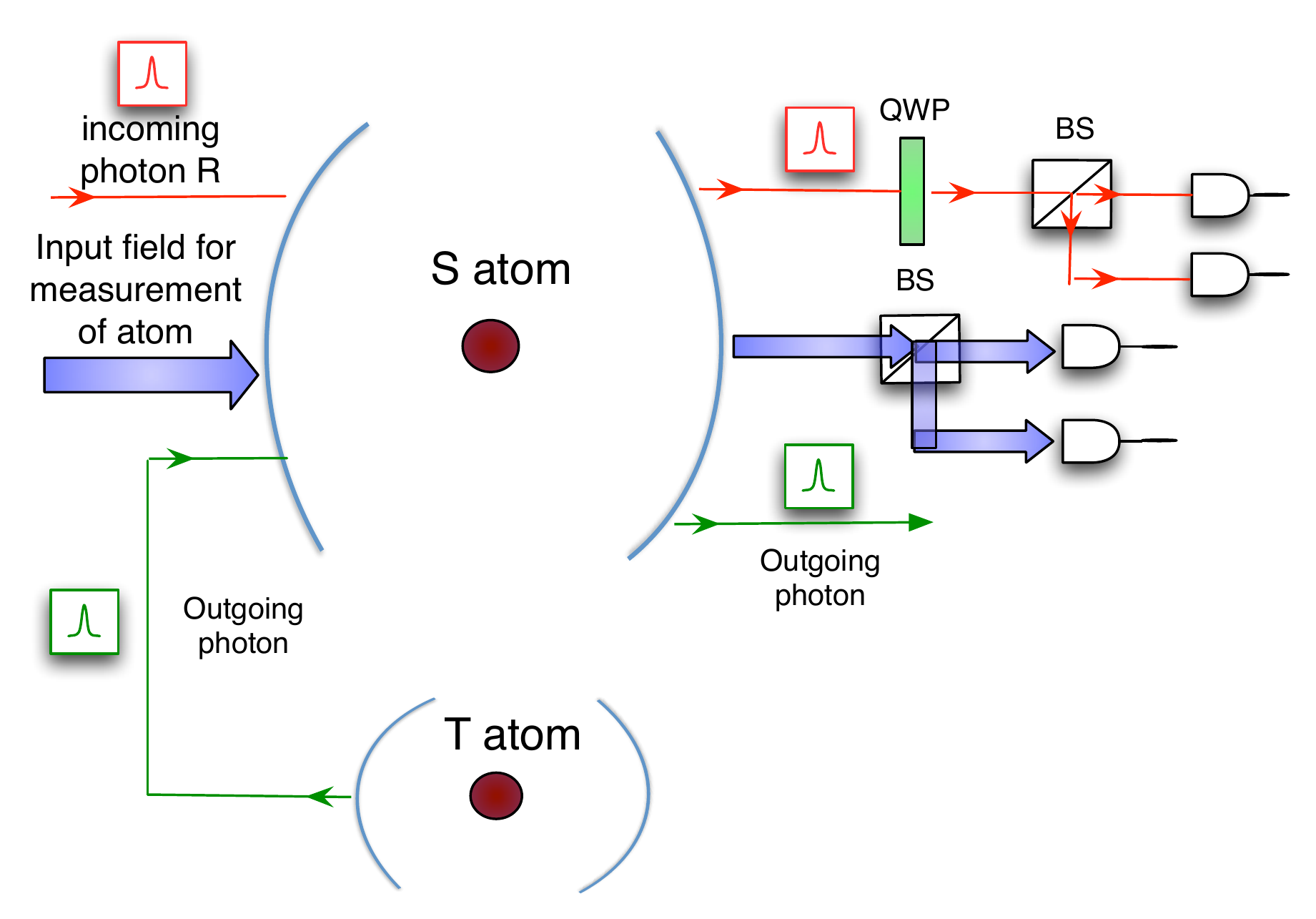}
\caption[fig:cavity]{(color online). A schematic view of the implementation
of the TEC protocol between a single atom and a single photon. There two
high Finesse cavities with $S$ atom and $T$ atom and the outgoing photon
from the cavity containing the $T$ atom enters the cavity containing $S$
atom. In addition, the the incoming photon from the previous repeater
station and the input field to control and measure the atom in the cavity
enters the cavity. At the output, an additional quarter wave plate (QWP) is
added for the X measurement of the incoming photon. The $Z$ measurement is
carried out by taking the photon through a beam splitter (BS) and detection.
}
\label{fig:cavity}
\end{figure}

For qubit $R_{i,j}$, the error channel can be characterized by the following
super-operator:
\begin{eqnarray}
\rho _{R}^{\prime } &=&\mathcal{E}_{R}\left( \rho _{RS}\right) =\eta \left(
1-\epsilon _{d}\right) \rho _{R}+\frac{\eta \epsilon _{d}}{4}%
\sum_{k=0}^{3}\sigma _{k}\rho \sigma _{k}  \notag \\
&&+\left( 1-\eta \right) \left\vert \text{vac}\right\rangle \left\langle
\text{vac}\right\vert ,
\end{eqnarray}%
where $\eta =\left( 1-p_{c}\right) e^{-L_{0}/L_{att}}$ is the transmission
probability, $1-\eta $ is the photon loss error probability, $\epsilon _{d}$
is the probability of depolarization error for a transmitted qubit, $%
\epsilon $ is the effective qubit error which takes into account-
measurement error $\epsilon _{m}$ and the gate error $\epsilon _{g}$,
respectively. Similarly, for the qubit $S_{i,j}$, the error channel can be
characterized as
\begin{equation}
\rho _{S}^{\prime }=\mathcal{E}_{S}\left( \rho _{RS}\right) =\left(
1-\epsilon _{d}\right) \rho _{S}+\frac{\epsilon _{d}}{4}\sum_{k=0}^{3}\sigma
_{k}\rho \sigma _{k},
\end{equation}%
which only has depolarization error but no photon loss error because the $S$
block consists of local qubits with no transmission loss. After the encoded
Bell measurement (with transversal CNOT gates), the errors of the two
encoding blocks become correlated. Hence, the measurement outcomes will
become correlated between the two blocks, in particular the measurement
outcome of the qubit pair $\left( R_{i,j},\,S_{i,j}\right) $ will become
correlated. For example, a $Y_{R}$ error on $R_{i,j}$ error will also induce
a correlated error $Y_{R}X_{S}$ on the qubit pair $\left(
R_{i,j},\,S_{i,j}\right) $. In order to compute the full distribution for
the measurement outcome at the encoded level, we take the following three
steps:

\begin{enumerate}
\item At the physical qubit level, we consider the errors associated with
the qubit-pair measurement $\left( X_{i,j}^{R},\,Z_{i,j}^{S}\right) $ for $%
i=1,\cdots ,n$ and $j=1,\cdots ,m$.

\item At the intermediate encoded level, we consider the errors associated
with the row-pair measurement $\left( X_{i}^{*R},Z_{i}^{*S}\right) $ for $%
i=1,\cdots ,n$.

\item At the logical encoded level, we consider the errors associated with
the encoded-pair measurement $\left( \tilde{X}^{R},\,\tilde{Z}^{S}\right) $.
\end{enumerate}

In the following, we will compute three error probability distributions
associated with these three different levels.

\subsection{Probability distribution for qubit-pair measurement}

First, we consider probability distribution associated with the qubit-pair
measurement $\left( X_{i,j}^{R},\,Z_{i,j}^{S}\right) $. For the ideal case
with no loss or operational errors, we will have outcomes $\left(
X_{i,j}^{R},Z_{i,j}^{S}\right) =\left( r_{i,j},s_{i,j}\right) $ with $%
r_{i,j},s_{i,j}=\pm 1$, but in the presence of errors the outcome will be $%
\left( X_{i,j}^{R},Z_{i,j}^{S}\right) =\left( \alpha r_{i,j},\beta
s_{i,j}\right) $ with $\left( \alpha ,\beta \right) =\left( 0,0\right)
,\left( \pm 1,\pm 1\right) ,\left( \pm 1,\mp 1\right) $, which corresponds
to the following cases:

\begin{enumerate}
\item $\left( \alpha ,\beta \right) =\left( 0,0\right) $: Erasure error on $%
R_{i,j}$ with probability $\epsilon _{e}=\Pr \left[
X_{i,j}^{R}=0,Z_{i,j}^{S}=0\right] =1-\eta $.

\item $\left( \alpha ,\beta \right) =\left( +1,-1\right) $: Spin-flip error
in outcomes $\left( r_{i,j},-s_{i,j}\right) $, with probability $\epsilon
_{X}=\Pr \left[ X_{i,j}^{R}=r_{i,j},Z_{i,j}^{S}=-s_{i,j}\right] =\eta \left(
\frac{1}{2}\epsilon _{d}+\frac{1}{4}\epsilon _{g}+\epsilon _{m}+O(\epsilon
_{d,g,m}^{2})\right) =\frac{1}{2}\eta\epsilon $, where the effective error
probability $\epsilon $ is defined as
\begin{equation}
\epsilon =\epsilon _{d}+\frac{1}{2}\epsilon _{g}+2\epsilon _{m}+O(\epsilon
_{d,g,m}^{2}).
\end{equation}

\item $\left( \alpha ,\beta \right) =\left( -1,-1\right) $:
Spin-\&-Phase-flip errors in outcomes $\left( -r_{i,j},-s_{i,j}\right) $,
with probability $\epsilon _{Y}=\Pr \left[
X_{i,j}^{R}=-r_{i,j},Z_{i,j}^{S}=-s_{i,j}\right] =\eta \left( \frac{1}{2}%
\epsilon _{d}+\frac{1}{4}\epsilon _{g}+\epsilon _{m}+O(\epsilon
_{d,g,m}^{2})\right) \approx \frac{1}{2}\eta \epsilon $, where the last step
upper bounds the probability of the case of $\left( \alpha ,\beta \right)
=\left( -1,-1\right) $.

\item $\left( \alpha ,\beta \right) =\left( -1,+1\right) $: Phase-flip error
in outcomes $\left( -r_{i,j},+s_{i,j}\right) $, with probability $\epsilon
_{Z}=\Pr \left[ X_{i,j}^{R}=-r_{i,j},Z_{i,j}^{S}=s_{i,j}\right] =\eta \left(
\frac{1}{2}\epsilon _{d}+\frac{1}{4}\epsilon _{g}+\epsilon _{m}+O(\epsilon
_{d,g,m}^{2})\right) =\frac{1}{2}\eta\epsilon $.

\item $\left( \alpha ,\beta \right) =\left( +1,+1\right) $: No change in the
measurement outcomes of $\left( r_{i,j},s_{i,j}\right) $, with probability $%
\epsilon _{I}=\Pr \left[ X_{i,j}^{R}=r_{i,j},Z_{i,j}^{S}=s_{i,j}\right]
=\eta \left( 1-\frac{3}{2}\epsilon _{d}-\frac{3}{4}\epsilon _{g}-3\epsilon
_{m}-O({\epsilon }_{d,g,m}^{2})\right) \approx \eta (1-\frac{3}{2}\epsilon )$%
.
\end{enumerate}

Note $\epsilon _{Y}\neq \left( \epsilon _{X}+\epsilon _{Y}\right) \left(
\epsilon _{Z}+\epsilon _{Y}\right) $ characterizing correlated errors. The
sum of these probabilities is unity, $\epsilon _{e}+\epsilon _{I}+\epsilon
_{X}+\epsilon _{Y}+\epsilon _{Z}=1$.

\subsection{Probability distribution for row-pair measurement}

We now consider the distribution associated with the row-pair measurement $%
\left( X_{i}^{\ast R},Z_{i}^{\ast S}\right) $. Suppose the ideal case, we
will have outcomes $\left( X_{i}^{\ast R},Z_{i}^{\ast S}\right) =\left(
r_{i},s_{i}\right) $ with $r_{i},s_{i}=0,\pm 1$, but in the presence of
errors the outcome will be $\left( X_{i}^{\ast R},Z_{i}^{\ast S}\right)
=\left( \alpha r_{i},\beta s_{i}\right) $ with $\left( \alpha ,\beta \right)
=\left( 0,\pm 1\right) \otimes \left( 0,\pm 1\right) $, with the following
probability distribution:
\begin{equation}
q_{\alpha ,\beta }:\equiv \Pr \left[ X_{i}^{\ast }=\alpha r_{i},Z_{i}^{\ast
}=\beta s_{i}\right] .
\end{equation}%
Note that $q_{\alpha ,,\beta }$ does not depend on the row index $i$,
because all rows have the same probability distribution.

In the measurement associated with the row-pair measurement, we may sum over
all possible error patterns of qubit-pair measurement, with $a$ photon loss
errors ($\epsilon _{e}$), $b$ spin-flip errors ($\epsilon _{X}$), $c$
spin-\&-phase-flip errors ($\epsilon _{Y}$), $d$ phase-flip errors ($%
\epsilon _{Z}$), and $e=m-a-b-c-d$ faithful measurements ($\epsilon _{I}$).

Because $X_{i}^{\ast R}=\prod_{j=1}^{m}X_{i,j}^{R}$ and $Z_{i}^{\ast S}=%
\text{sign}\left[ \sum_{j=1}^{n}Z_{i,j}\right] $, the conditions are
\begin{equation}
\alpha =\left\{
\begin{tabular}{ll}
$0$ & if $a \geq 1$ \\
$+1$ & if $a =0$ \& $\left( c+d\right) $ even \\
$-1$ & if $a =0$ \& $\left( c+d\right) $ odd%
\end{tabular}%
\right.
\end{equation}%
and%
\begin{equation}
\beta =\left\{
\begin{tabular}{ll}
$0$ & if $2\left( b+c\right) =m-a$ \\
$+1$ & if $2\left( b+c\right) <m-a$ \\
$-1$ & if $2\left( b+c\right) >m-a$%
\end{tabular}%
\right. .
\end{equation}

Then we can compute all $\left\{ q_{\alpha ,\beta }\right\} $. For example,
\begin{eqnarray*}
&&q_{0,0} \\
&=&\sum_{a,b,c,d}^{m}\delta _{a\geq 1}\delta _{2\left( b+c\right)
=m-a}\left(
\begin{array}{c}
m \\
a,b,c,d%
\end{array}%
\right) \epsilon _{e}^{a}\epsilon _{X}^{b}\epsilon _{Y}^{c}\epsilon
_{Z}^{d}\epsilon _{I}^{m-a-b-c-d} \\
&=&\sum_{a=1}^{m}\,\,\sum_{2w=m-a}\left(
\begin{array}{c}
m \\
a,w%
\end{array}%
\right) \epsilon _{e}^{a}\left( \epsilon _{X}+\epsilon _{Y}\right)
^{w}\left( \epsilon _{Z}+\epsilon _{I}\right) ^{m-a-w}
\end{eqnarray*}%
with
\begin{equation*}
\delta _{\mathrm{cond}}:\equiv \left\{
\begin{tabular}{ll}
$1$ & if $\mathrm{cond=true}$ \\
$0$ & if $\mathrm{cond=false}$%
\end{tabular}%
\right.
\end{equation*}%
and multinomial%
\begin{eqnarray*}
&&\left(
\begin{array}{c}
m \\
a,b,c,d%
\end{array}%
\right) \\
&=&\left(
\begin{array}{c}
m \\
a%
\end{array}%
\right) \left(
\begin{array}{c}
m-a \\
b%
\end{array}%
\right) \left(
\begin{array}{c}
m-a-b \\
c%
\end{array}%
\right) \left(
\begin{array}{c}
m-a-b-c \\
d%
\end{array}%
\right) .
\end{eqnarray*}

Similarly we can compute the remaining $q_{\alpha ,\beta }$:
\begin{equation*}
q_{0,+1}=\sum_{a=1}^{m}\,\,\sum_{2w<m-a}\left(
\begin{array}{c}
m \\
a,w%
\end{array}%
\right) \epsilon _{e}^{a}\left( \epsilon _{X}+\epsilon _{Y}\right)
^{w}\left( \epsilon _{Z}+\epsilon _{I}\right) ^{m-a-w}
\end{equation*}%
\begin{equation*}
q_{0,-1}=\sum_{a=1}^{m}\,\,\sum_{2w>m-a}\left(
\begin{array}{c}
m \\
a,w%
\end{array}%
\right) \epsilon _{e}^{a}\left( \epsilon _{X}+\epsilon _{Y}\right)
^{w}\left( \epsilon _{Z}+\epsilon _{I}\right) ^{m-a-w}.
\end{equation*}%
For the case with no photon losses, we have
\begin{eqnarray*}
q_{+1,0} &=&\sum_{b,c,d}^{m}\delta _{\left( c+d\right) \mathrm{even}}\delta
_{2\left( b+c\right) =m}\left(
\begin{array}{c}
m \\
b,c,d%
\end{array}%
\right) \epsilon _{X}^{b}\epsilon _{Y}^{c}\epsilon _{Z}^{d}\epsilon
_{I}^{m-b-c-d} \\
&=&\sum_{v\,\mathrm{even}}\,\,\sum_{c=0}^{v}\,\,\sum_{b=\frac{m}{2}-c}\left(
\begin{array}{c}
m \\
v-c,c,b%
\end{array}%
\right) \epsilon _{X}^{b}\epsilon _{Y}^{c}\epsilon _{Z}^{v-c}\epsilon
_{I}^{m-b-v}
\end{eqnarray*}%
\begin{equation*}
q_{+1,+1}=\sum_{v\,\mathrm{even}}\,\,\sum_{c=0}^{v}\,\,\sum_{b<\frac{m}{2}%
-c}\left(
\begin{array}{c}
m \\
v-c,c,b%
\end{array}%
\right) \epsilon _{X}^{b}\epsilon _{Y}^{c}\epsilon _{Z}^{v-c}\epsilon
_{I}^{m-b-v}
\end{equation*}%
\begin{equation*}
q_{+1,-1}=\sum_{v\,\mathrm{even}}\,\,\sum_{c=0}^{v}\,\,\sum_{b>\frac{m}{2}%
-c}\left(
\begin{array}{c}
m \\
v-c,c,b%
\end{array}%
\right) \epsilon _{X}^{b}\epsilon _{Y}^{c}\epsilon _{Z}^{v-c}\epsilon
_{I}^{m-b-v},
\end{equation*}%
and
\begin{equation*}
q_{-1,0}=\sum_{v\,\mathrm{odd}}\,\,\sum_{c=0}^{v}\,\,\sum_{b=\frac{m}{2}%
-c}\left(
\begin{array}{c}
m \\
v-c,c,b%
\end{array}%
\right) \epsilon _{X}^{b}\epsilon _{Y}^{c}\epsilon _{Z}^{v-c}\epsilon
_{I}^{m-b-v}
\end{equation*}%
\begin{equation*}
q_{-1,+1}=\sum_{v\,\mathrm{odd}}\,\,\sum_{c=0}^{v}\,\,\sum_{b<\frac{m}{2}%
-c}\left(
\begin{array}{c}
m \\
v-c,c,b%
\end{array}%
\right) \epsilon _{X}^{b}\epsilon _{Y}^{c}\epsilon _{Z}^{v-c}\epsilon
_{I}^{m-b-v}
\end{equation*}%
\begin{equation*}
q_{-1,-1}=\sum_{v\,\mathrm{odd}}\,\,\sum_{c=0}^{v}\,\,\sum_{b>\frac{m}{2}%
-c}\left(
\begin{array}{c}
m \\
v-c,c,b%
\end{array}%
\right) \epsilon _{X}^{b}\epsilon _{Y}^{c}\epsilon _{Z}^{v-c}\epsilon
_{I}^{m-b-v}.
\end{equation*}

\subsection{Probability distribution for encoded-pair measurement}

We now consider the distribution associated with the encoded-pair
measurement $\left( \tilde{X},\tilde{Z}\right) $. For the ideal case, we
will have outcomes $\left( \tilde{M}_{X}^{R},\tilde{M}_{Z}^{S}\right)
=\left( \tilde{X}^{R},\tilde{Z}^{S}\right) $ with $\tilde{X}^{R},\tilde{Z}%
^{S}=\pm 1$, but in the presence of errors the outcome will be $\left(
\tilde{M}_{X}^{R},\tilde{M}_{Z}^{S}\right) =\left( \alpha \tilde{X}%
^{R},\beta \tilde{Z}^{S}\right) $ with $\left( \alpha ,\beta \right) =\left(
0,\pm 1\right) \otimes \left( 0,\pm 1\right) $, with the following
probability distribution:
\begin{equation}
p_{\alpha ,\beta }:\equiv \Pr \left[ \tilde{M}_{X}=\alpha \tilde{X}^{R},%
\tilde{M}_{Z}=\beta \tilde{Z}^{S}\right] .
\end{equation}%
In the measurement associated with the encoded-pair measurement, we may sum
over all possible error patterns of row-pair measurements, with $a$
instances of ($q_{0,0}$), $b$ instances of ($q_{1,0}$), $c$ instances of ($%
q_{-1,0}$), $d$ instances ($q_{0,1}$), $e$ instances of ($q_{1,1}$), $f$
instances of ($q_{-1,1}$), $g$ instances of ($q_{0,-1}$), $h$ instances of ($%
q_{1,-1}$), $i$ instances of ($q_{-1,-1}$), with $a+b+c+d+e+f+g+h+i=n$.

Because $\tilde{M}_{X}^{R}=\text{sign}\left[ \sum_{i=1}^{n}X_{i}^{\ast }%
\right] $and $\tilde{M}_{Z}^{S}=\prod_{i=1}^{n}Z_{i}^{\ast }$, then the
conditions are
\begin{equation}
\tilde{M}_{X}^{R}=\left\{
\begin{tabular}{ll}
$0$ & if $2\left( b+e+h\right) =n-\left( a+d+g\right) $ \\
$+1$ & if $2\left( b+e+h\right) >n-\left( a+d+g\right) $ \\
$-1$ & if $2\left( b+e+h\right) <n-\left( a+d+g\right) $%
\end{tabular}%
\right.
\end{equation}%
and
\begin{equation}
\tilde{M}_{Z}^{S}=\left\{
\begin{tabular}{ll}
$0$ & if $a+b+c\geq 1$ \\
$+1$ & if $a+b+c=0$ \& $\left( g+h+i\right) $ even \\
$-1$ & if $a+b+c=0$ \& $\left( g+h+i\right) $ odd%
\end{tabular}%
\right. .
\end{equation}%
Finally, we can compute all $\left\{ p_{\alpha ,\beta }\right\} $:

\begin{widetext}

\begin{eqnarray*}
p_{0,0} &=&\sum_{a,b,c,d,e,f,g,h}\delta _{a+b+c\geq 1}\delta _{2\left(
b+e+h\right) =n-\left( a+d+g\right) }\left(
\begin{array}{c}
n \\
a,b,c,d,e,f,g,h%
\end{array}%
\right)
q_{0,0}^{a}q_{1,0}^{b}q_{-1,0}^{c}q_{0,1}^{d}q_{1,1}^{e}q_{-1,1}^{f}q_{0,-1}^{g}q_{1,-1}^{h}q_{-1,-1}^{n-\left( a+b+c+d+e+f+g+h\right) }
\\
&=&\sum_{a,b,c,v,w}\delta _{a+b+c\geq 1}\delta _{2\left( b+w\right)
=n-a-v}\left(
\begin{array}{c}
n \\
a,b,c,v,w%
\end{array}%
\right) q_{0,0}^{a}q_{1,0}^{b}q_{-1,0}^{c}\left( q_{0,1}+q_{0,-1}\right)
^{v}\left( q_{1,1}+q_{-1,1}\right) ^{w}\left( q_{1,-1}+q_{-1,-1}\right)
^{n-\left( a+b+c+v+w\right) }
\end{eqnarray*}%
\[
p_{1,0}=\sum_{a,b,c,v,w}\delta _{a+b+c\geq 1}\delta _{2\left( b+w\right)
>n-a-v}\left(
\begin{array}{c}
n \\
a,b,c,v,w%
\end{array}%
\right) q_{0,0}^{a}q_{1,0}^{b}q_{-1,0}^{c}\left( q_{0,1}+q_{0,-1}\right)
^{v}\left( q_{1,1}+q_{-1,1}\right) ^{w}\left( q_{1,-1}+q_{-1,-1}\right)
^{n-\left( a+b+c+v+w\right) }
\]%
\[
p_{-1,0}=\sum_{a,b,c,v,w}\delta _{a+b+c\geq 1}\delta _{2\left( b+w\right)
<n-a-v}\left(
\begin{array}{c}
n \\
a,b,c,v,w%
\end{array}%
\right) q_{0,0}^{a}q_{1,0}^{b}q_{-1,0}^{c}\left( q_{0,1}+q_{0,-1}\right)
^{v}\left( q_{1,1}+q_{-1,1}\right) ^{w}\left( q_{1,-1}+q_{-1,-1}\right)
^{n-\left( a+b+c+v+w\right) }
\]%
\[
p_{0,1}=\sum_{d,e,g,h,i}\delta _{\left( g+h+i\right) \,\mathrm{even}}\delta
_{2\left( e+h\right) =n-d-g}\left(
\begin{array}{c}
n \\
d,e,g,h,i%
\end{array}%
\right) q_{0,1}^{d}q_{1,1}^{e}q_{-1,1}^{n-\left( d+e+g+h+i\right)
}q_{0,-1}^{g}q_{1,-1}^{h}q_{-1,-1}^{i}
\]%
\[
p_{1,1}=\sum_{d,e,g,h,i}\delta _{\left( g+h+i\right) \,\mathrm{even}}\delta
_{2\left( e+h\right) >n-d-g}\left(
\begin{array}{c}
n \\
d,e,g,h,i%
\end{array}%
\right) q_{0,1}^{d}q_{1,1}^{e}q_{-1,1}^{n-\left( d+e+g+h+i\right)
}q_{0,-1}^{g}q_{1,-1}^{h}q_{-1,-1}^{i}
\]%
\[
p_{-1,1}=\sum_{d,e,g,h,i}\delta _{\left( g+h+i\right) \,\mathrm{even}}\delta
_{2\left( e+h\right) <n-d-g}\left(
\begin{array}{c}
n \\
d,e,g,h,i%
\end{array}%
\right) q_{0,1}^{d}q_{1,1}^{e}q_{-1,1}^{n-\left( d+e+g+h+i\right)
}q_{0,-1}^{g}q_{1,-1}^{h}q_{-1,-1}^{i}
\]%
\[
p_{0,-1}=\sum_{d,e,g,h,i}\delta _{\left( g+h+i\right) \,\mathrm{odd}}\delta
_{2\left( e+h\right) =n-d-g}\left(
\begin{array}{c}
n \\
d,e,g,h,i%
\end{array}%
\right) q_{0,1}^{d}q_{1,1}^{e}q_{-1,1}^{n-\left( d+e+g+h+i\right)
}q_{0,-1}^{g}q_{1,-1}^{h}q_{-1,-1}^{i}
\]%
\[
p_{1,-1}=\sum_{d,e,g,h,i}\delta _{\left( g+h+i\right) \,\mathrm{odd}}\delta
_{2\left( e+h\right) >n-d-g}\left(
\begin{array}{c}
n \\
d,e,g,h,i%
\end{array}%
\right) q_{0,1}^{d}q_{1,1}^{e}q_{-1,1}^{n-\left( d+e+g+h+i\right)
}q_{0,-1}^{g}q_{1,-1}^{h}q_{-1,-1}^{i}
\]%
\[
p_{-1,-1}=\sum_{d,e,g,h,i}\delta _{\left( g+h+i\right) \,\mathrm{odd}}\delta
_{2\left( e+h\right) <n-d-g}\left(
\begin{array}{c}
n \\
d,e,g,h,i%
\end{array}%
\right) q_{0,1}^{d}q_{1,1}^{e}q_{-1,1}^{n-\left( d+e+g+h+i\right)
}q_{0,-1}^{g}q_{1,-1}^{h}q_{-1,-1}^{i}.
\]%
\end{widetext}

\section{Overhead from fault-tolerant state preparation}

The qubit overhead and the time overhead are closely related for
fault-tolerant state preparation. To understand this better, consider the
syndrome measurement of the stabilizers $S_{i,0}$. The stabilizers can be
measured in parallel with two time steps. For instance, suppose we have a $%
(4,4)$ QPC, we need one time step to measure the rows $\{1,2\}$ and rows $%
\{3,4\}$ simultaneously and another time step to measure the rows $\{2,3\}$.
To achieve this, we need to prepare two GHZ states of $8$ qubits each
simultaneosuly. So, it takes $16$ qubits in total to prepare one GHZ state
and $32$ qubits to prepare two GHZ states. It is fairly straigtforward that
it will take an additional overhead of $2mn$ qubits to measure the
stabilizers within two steps (i.e, after the creation of the GHZ states).
Similarly, one can also consider the overhead associated with the
fault-tolerant preparation of the GHZ state. Using this procedure it will
take an additional $4mn$ qubits for the fault-tolerant preparation of the
encoded Bell pair.\\

Suppose, if we can achieve very fast quantum gates with a high efficiency, then
we can further improve the overhead in the number of qubits by using the
same $2m$ qubits to recreate a GHZ state and to measure all the stabilizers
of the QPC. This can be achieved with a overhead of just $4m$ qubits, but
the time-overhead is scaled by a factor of $(n-1)$ compared to the previous
preparation scheme for the creation of an encoded EPR pair.

It is for this reason, the cost function introduced in the manuscript
considers only the qubits required for the creation of the encoded Bell pair
and does not consider the additional qubit overhead required for the
fault-tolerant preparation as there is more than one way to do so. But the
analysis of the cost function will be very similar to the one considered in
the manuscript. While we discussed a specific fault-tolerant preparation
scheme of Brooks and Preskill \cite{Brooks}, it will take future work to
determine the best fault-tolerant preparation scheme for the QPC given the
overhead in terms of qubits and time.
\color{black}
\section{Fault tolerant properties of QPC}

\begin{figure}[h]
\centering
\subfigure[]{
   \label{fig:thresh1}
  \includegraphics[width=8cm]{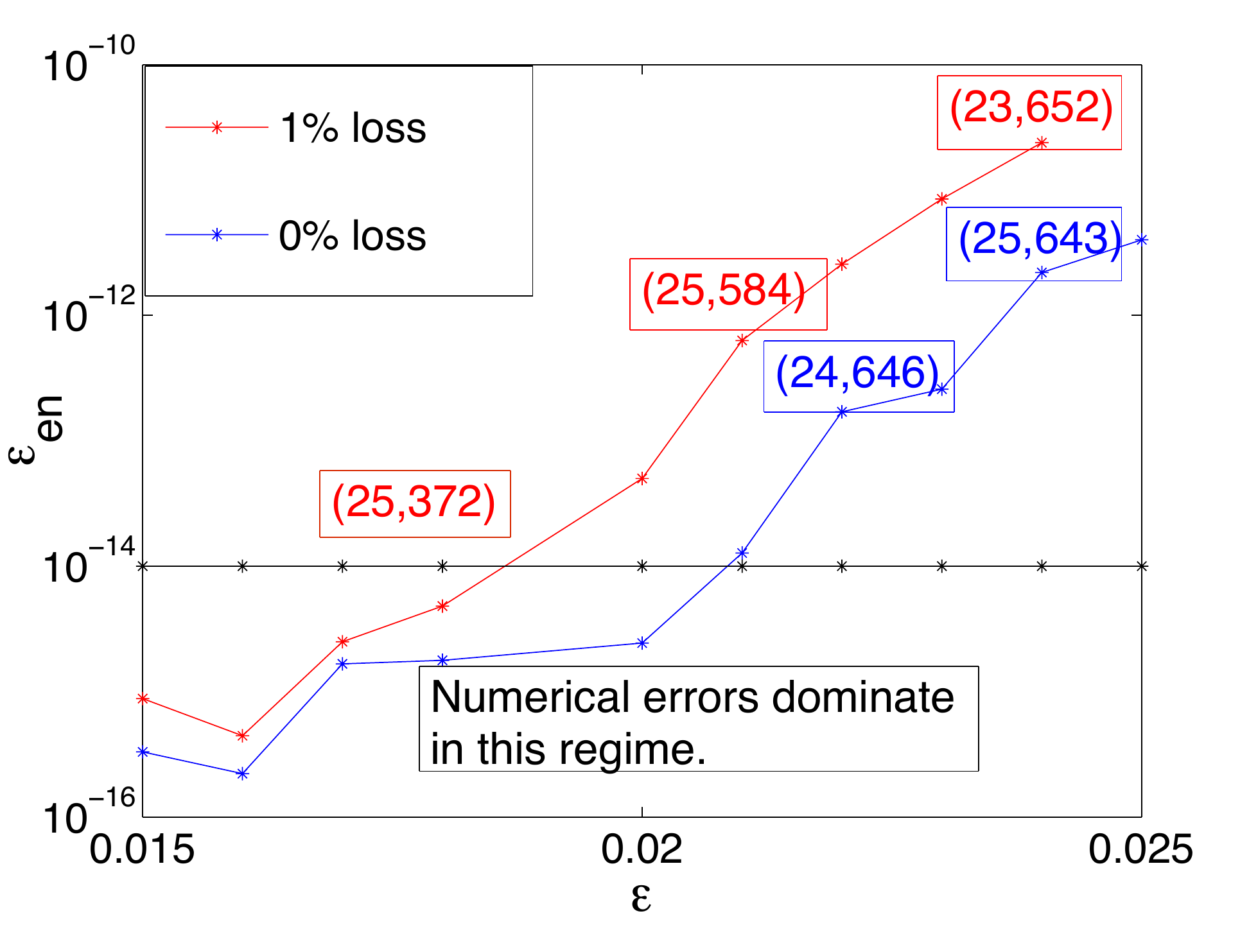}
 }
\subfigure[]{
   \label{fig:thresh1}
  \includegraphics[width=8cm]{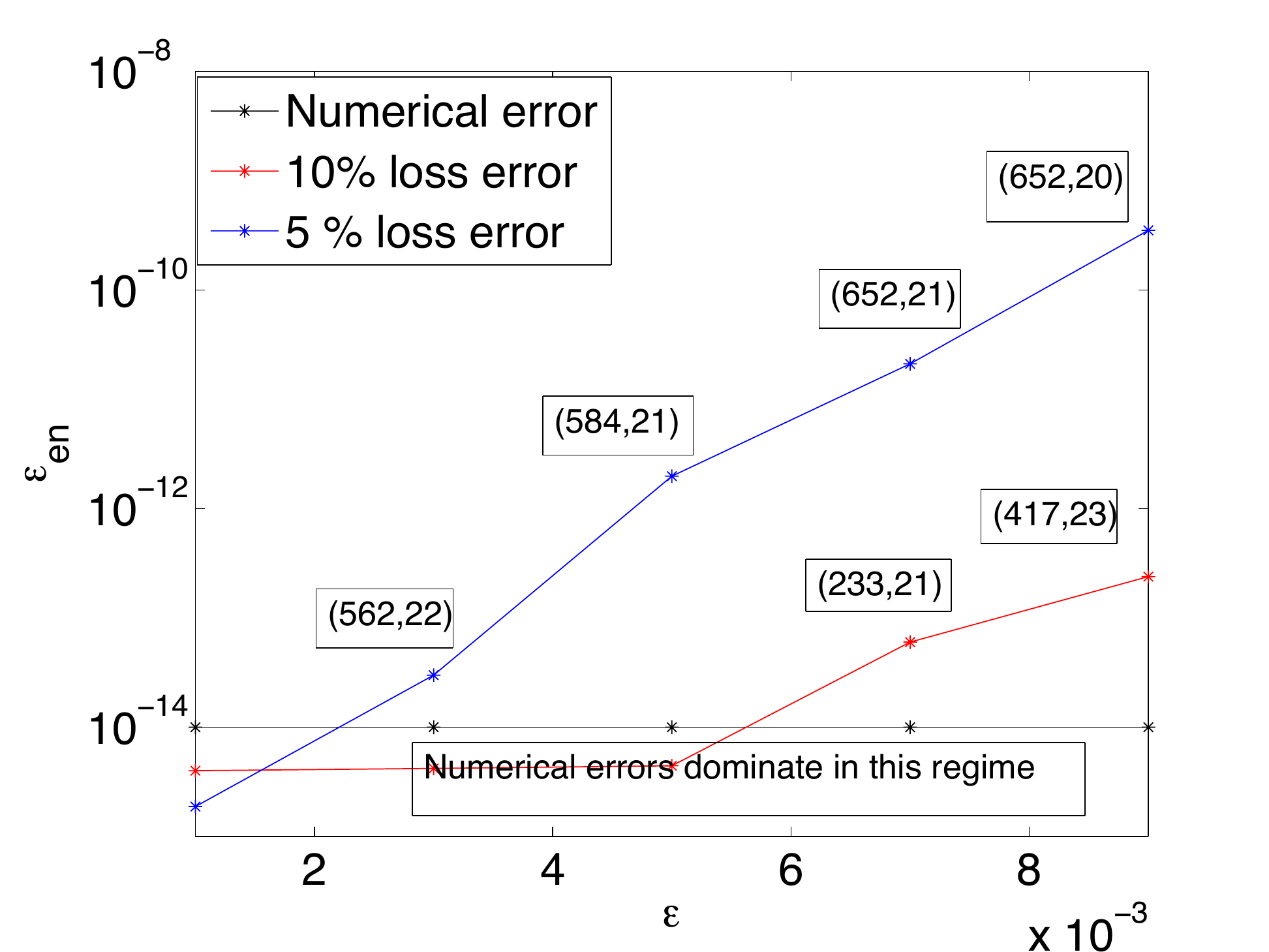}
 }
\caption[fig:fa1]{(color online). Optimum logical error rate $\protect%
\epsilon _{en}$ vs physical error rate $\protect\epsilon $ (a) for no losses
and $1\%$ loss errors. (b) for $5\%$ loss errors and $10\%$ loss errors.
Numerical errors begin to dominate in the range $\protect\epsilon %
_{en}\approx 10^{-14}$.}
\label{fig:fa1}
\end{figure}

An important difference between fault-tolerant quantum computers and
fault-tolerant quantum repeaters is that loss errors play an important role
in the latter. For a single transmission of QPC between neighboring QR
stations, we can define the effective encoded error rate to be $\epsilon
_{en}=(1-p_{1,1})$, which takes into account both heralded failure and
quantum bit error rates. Analogous to the recent study of Brooks and Preskill \cite{Brooks} on Bacon-Shor codes
\cite{Brooks}, we show in Fig.~\ref{fig:fa1} that it is possible to suppress
the encoded error to $\epsilon _{en}\approx 2 \times 10^{-14}$ by choosing an
appropriate encoding with a large number of qubits in a specific range for $%
\epsilon $, (a) $1.5\times 10^{-2}\leq \epsilon \leq 2.5\times 10^{-2}$ in
the absence of loss errors and in the presence of low loss errors $(1\%)$.
(b) $1\times 10^{-3}\leq \epsilon \leq 9\times 10^{-3}$ in the presence of
higher loss errors $(5\%,10\%)$.

Alternatively, we confirm with numerical calculations that in the absence of
loss errors and in the presence of loss errors (up to $(10\%)$), it is
possible to arbitrarily suppress the encoded error rate - which accounts for
both the bit-error rate and the failure probabilities to $\epsilon
_{en}\approx 10^{-14}$. Below $10^{-14}$, numerical errors begin to play an
important role. The results are summarized in the Table~\ref{tab:QPCcodes}.

\begin{table}[h]
\begin{center}
{\normalsize
\begin{tabular}{|c|c|c|c|c|}
\hline
$\epsilon \backslash {p_{c}}(1-\eta )$ & {$0\%$} & $1\%$ & $5\%$ & $10\%$ \\
\hline
$10^{-3}$ & {(19,13)} & (28,14) & (83,17) & (309,19) \\ \hline
$6\times 10^{-4}$ & {(17,11)} & (25,12) & (77,16) & (290,18) \\ \hline
$3\times 10^{-4}$ & {(13,11)} & (20,12) & (54,14) & (180,17) \\ \hline
$10^{-4}$ & {(11,9)} & (17,10) & (49,13) & (170,16) \\ \hline\hline
\end{tabular}
}
\end{center}
\caption[tab:QPCcodes]{QPC codes that are required to achieve an encoded
error rate of $\protect\epsilon _{en}\approx 2\times 10^{-14}$ for different
physical error rates $\protect\epsilon $ in the presence of varying losses
in $\%$. }
\label{tab:QPCcodes}
\end{table}

\section{Details of the optimization algorithm}

\begin{figure}[h]
\centering
\includegraphics[height=3.4in,width=3.2in,angle=0]{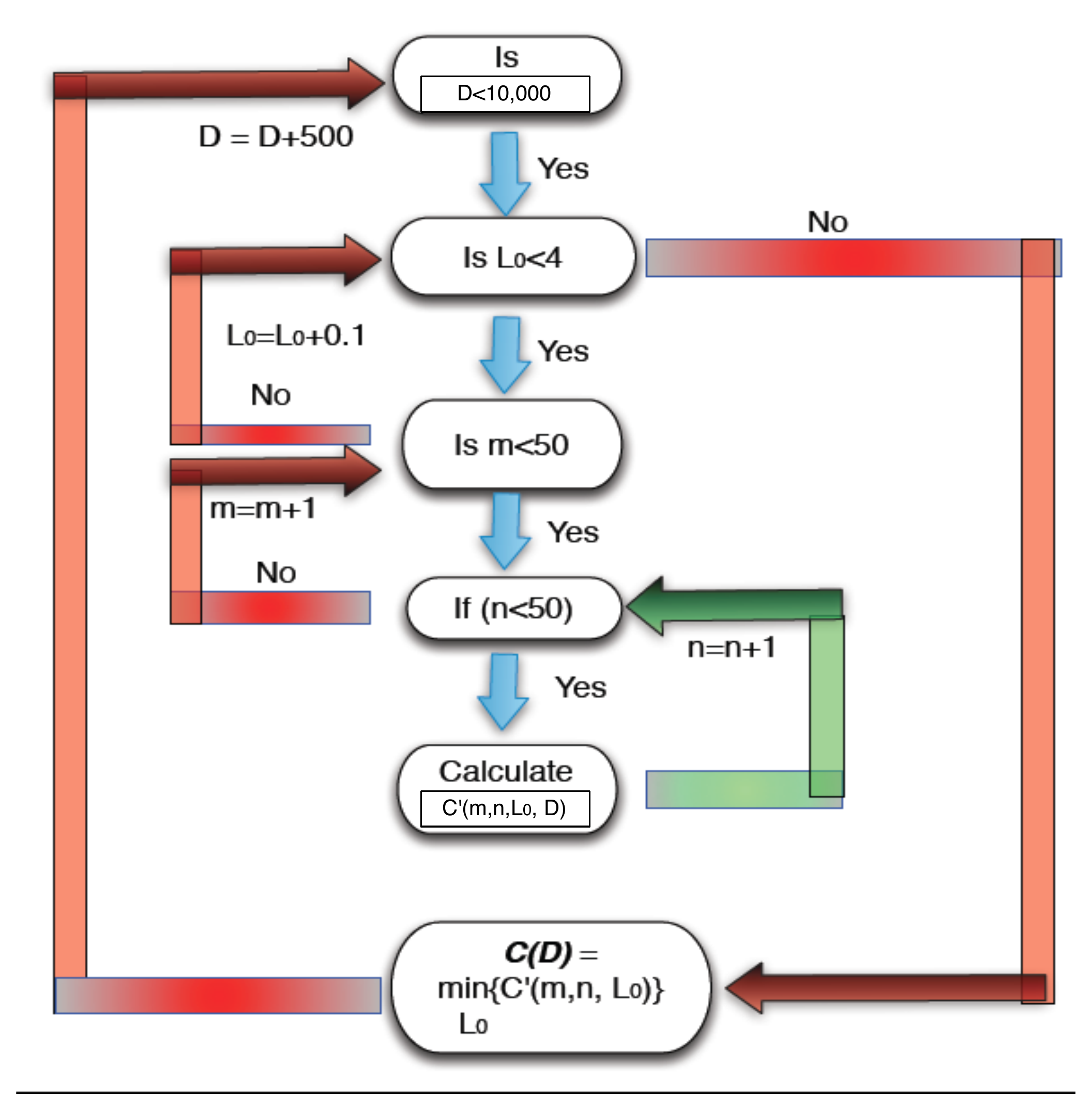}
\caption[fig:appendix2]{(color online). The flow chart of the algorithm to
find the optimized QR parameters. The units of $L_{tot}$ and $L_{0}$ are in $%
\mbox{km}$ and are ignored in the figure for convenience.}
\label{fig:appendix2}
\end{figure}

A self explanatory flow chart of the optimization algorithm used in the
Letter for the minimal cost coefficient of third class of QRs is shown in
Fig.~\ref{fig:appendix2}. We start the search with $L_{tot}=500\ mbox{km}$, $%
L_{0}=1\ \mbox{km}$ and $m=n=2$.

\section{Scaling of the cost coefficient}

\begin{figure}[t]
\centering
\subfigure[]{
   \label{fig:app7a}
  \includegraphics[width=4cm]{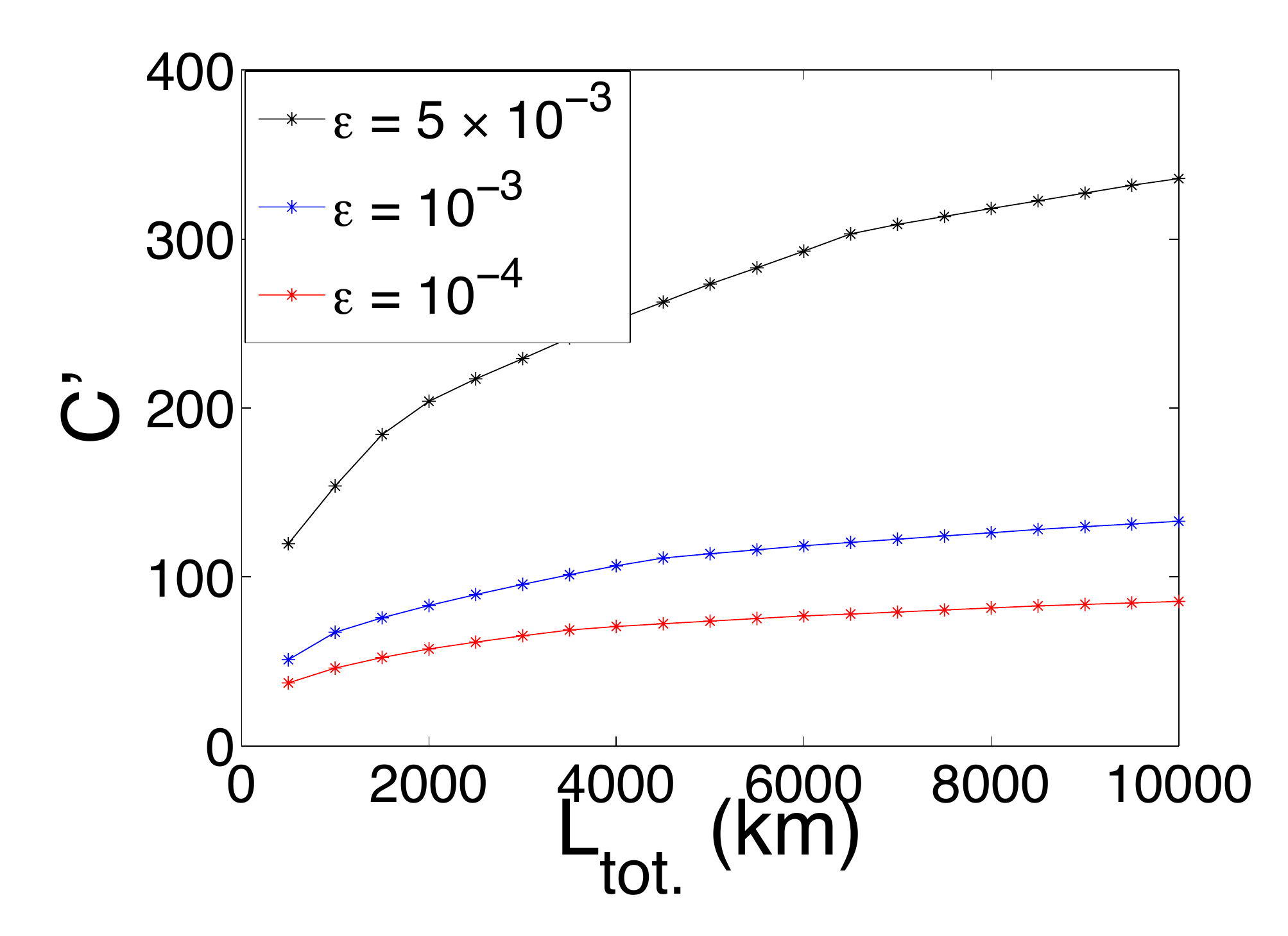}
 }
\subfigure[]{
   \label{fig:app7b}
 \includegraphics[width=4cm]{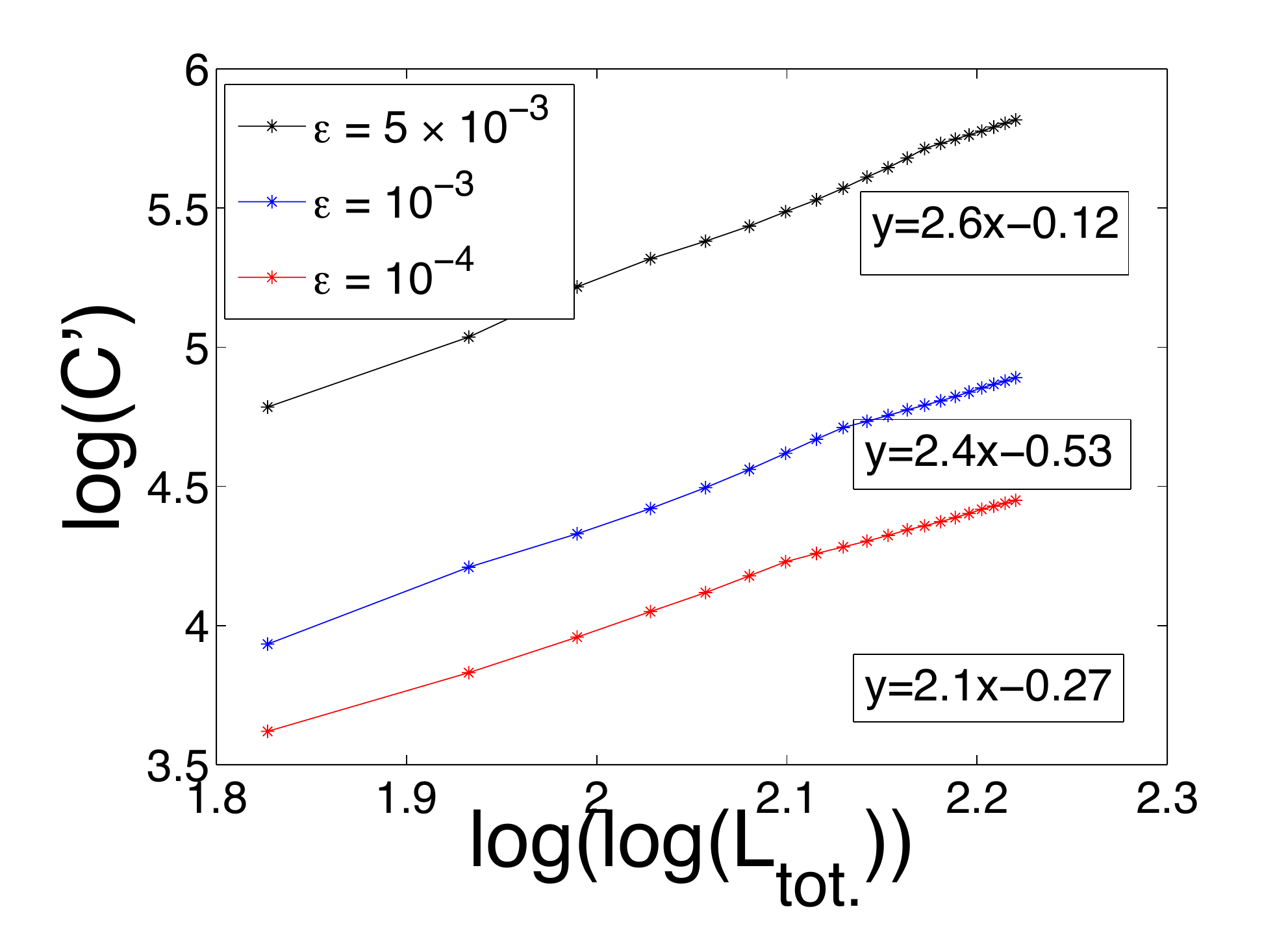}
 }
\subfigure[]{
   \label{fig:app7c}
  \includegraphics[width=4cm]{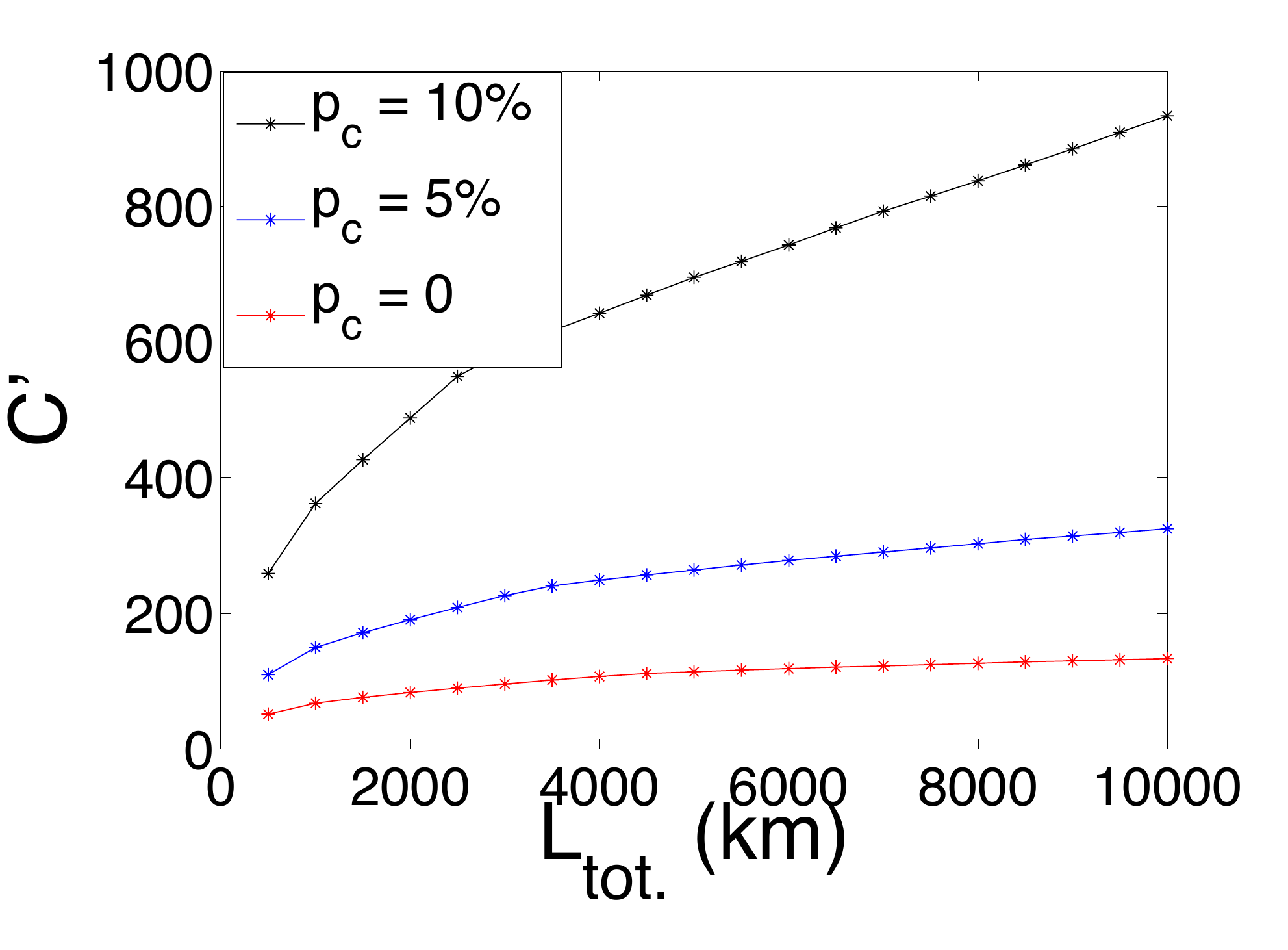}
 }
\subfigure[]{
   \label{fig:app7d}
  \includegraphics[width=4cm]{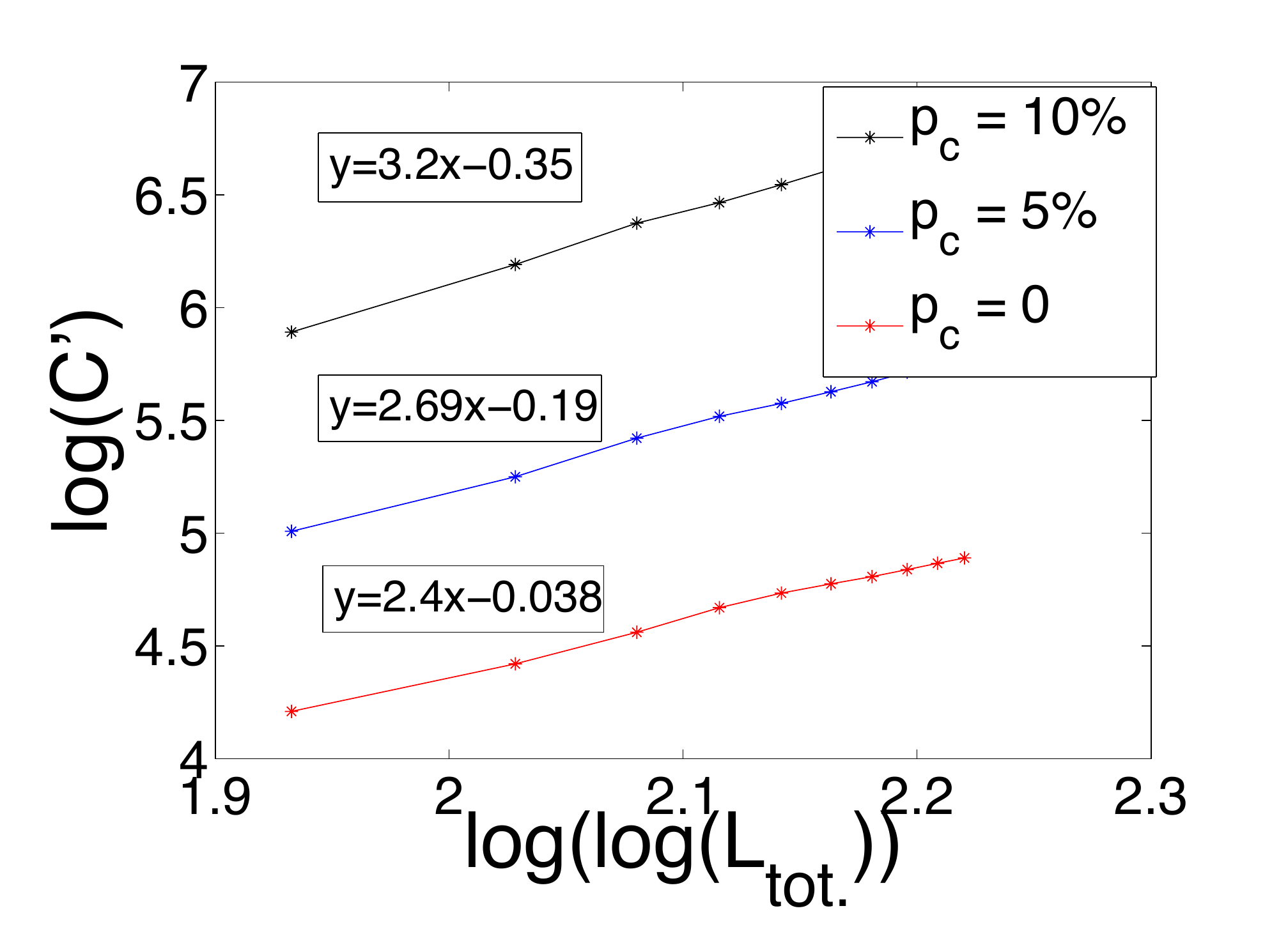}
 }
\caption[fig:appendix3]{ (color online). (a) $C^{\prime }(L_{tot})$ for
various $\protect\varepsilon $'s up to $10^{4}$ km, (b) Indication of
poly-logarithmic scaling of (a), (c) $C^{\prime }(L_{tot})$ for $\protect%
\varepsilon =10^{-3}$ up to $10^{4}$ km with coupling losses, (d) Indication
of poly-logarithmic scaling of (c). We assume $t_{0}=1$ in (a), (b), (c) and
(d) for convenience. }
\label{fig:appendix3}
\end{figure}

In the absence of a QR, the cost coefficient scales exponentially with the
distance across which the communication is desired. In the presence of our
QRs, a numerical investigation (Fig.~\ref{fig:appendix3}) of the cost
coefficient indicates that it has a poly-logarithmic scaling with the total
distance of communication up to $L_{tot}=10^{4}\ \mbox{km}$ in the absence and in
the presence of coupling losses (up to $pc=10\%$), respectively. In the
regime where $\varepsilon $ is smaller than $10^{-3}$ and there are no
coupling losses, the QBER and the success probability are dominated by the
photon loss errors and the cost coefficient scales as $\approx O(logD)^{2}$.
As the contribution of $\varepsilon $ to the final success probability and
QBER increases, the quadratic scaling breaks, but the scaling of the cost
coefficient still seems to be poly-logarithmic with distance.

\section{Generalized cost coefficient}

The cost coefficient introduced in the letter is defined for the case when
the cost of the qubits are expensive, but it is possible to envision a
scenario, where qubits may be cheap. Taking this into account, we can define
the \emph{generalized cost coefficient} to be

\begin{equation}
C^{\prime }= \frac{{(2nm)}^k}{R\cdot L_o},
\end{equation}
\newline
where $k$ is a constant satisfying $0 \leq k \leq 1$. The choice of the
above definition is guided by the constraint to obtain a unitless cost
coefficient which scales polynomial in the number of qubits.For $k=0$,
qubits cost absolutely nothing and $k=1$ corresponds to the case considered
in the letter, which takes into account the cost of the qubits. A comparison
of the cost coefficients for different $k$'s is shown in Fig. ~\ref%
{fig:costf3}.

\begin{figure}[h]
\centering
{\ \includegraphics[width=6cm]{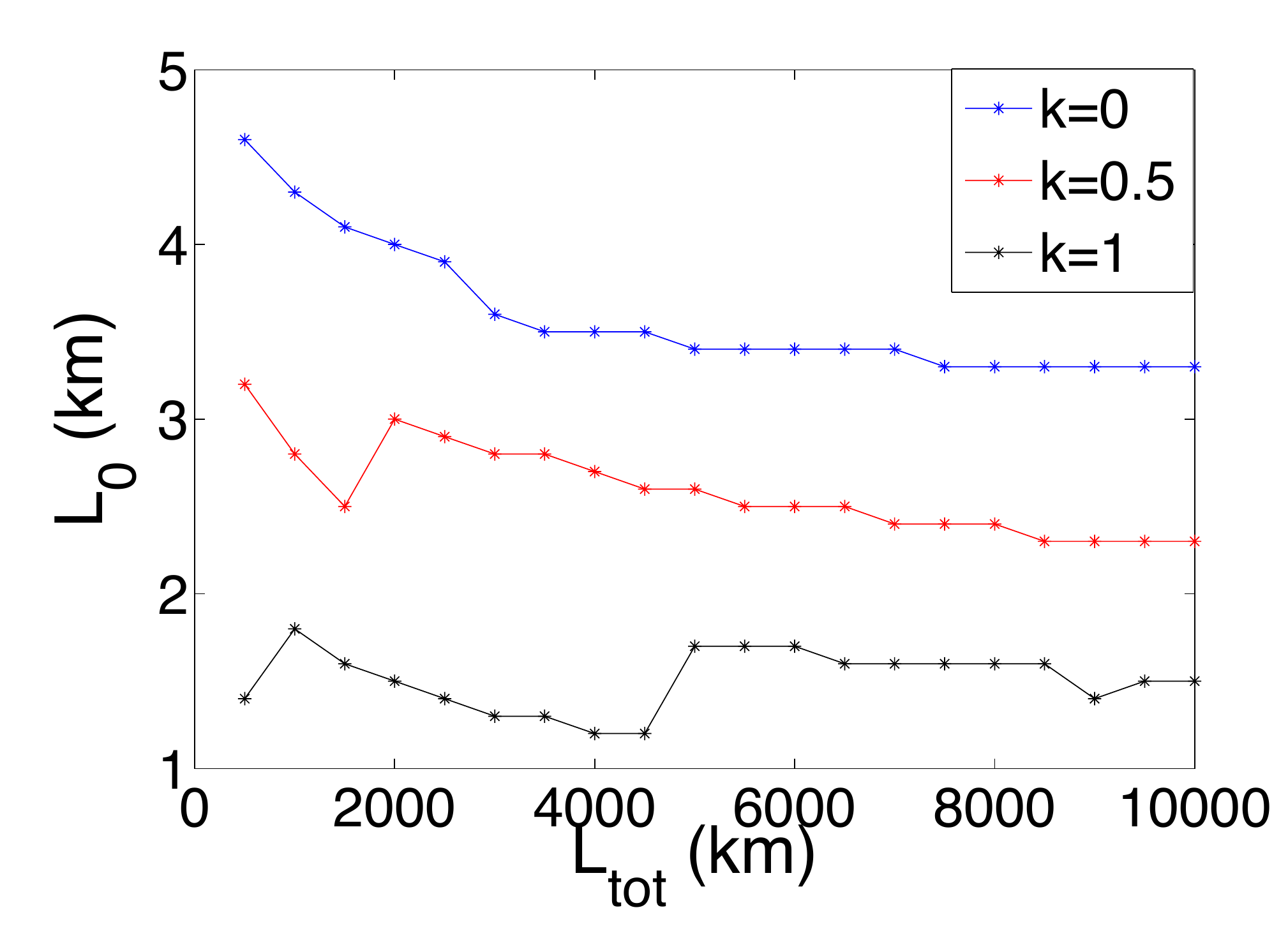} }
\caption[fig:costf3]{ (color online). For a range of search $2\leq \
(n,m)\leq \ 50$, $\protect\epsilon =10^{-3}$ and $p_{c}=0$, Optimal repeater
spacing for various $k$'s. Note that for $k=0$, it is possible to have a
larger repeater spacing by increasing the range of search. }
\label{fig:costf3}
\end{figure}

Interestingly, it is possible to have higher repeater spacings for the case
where the qubits are cheap as shown in Fig. ~\ref{fig:costf3}. For $k=0$,
one can have a higher repeater spacing by increasing the range of search. To
provide an estimate, for $\epsilon =10^{-3}$ and $p_{c}=0$, to generate a
secret key across $1000\ \mbox{km}$, with 800 qubits per repeater station,
one can have a repeater spacing of $4.3\ \mbox{km}$ and with $8500$ qubits
per repeater station, one can have a repeater spacing of $6.3\ \mbox{km}$.
Similarly, to generate a secret key across $10,000\ \mbox{km}$, with 1000
qubits per repeater station, one can have a repeater spacing of $4.1\ %
\mbox{km}$ and with $9100$ qubits per repeater station, one can have a
repeater spacing of $5.5\ \mbox{km}$.

\end{document}